\documentclass[fleqn,usenatbib]{mnras}
\usepackage{graphicx} 
\usepackage{newtxtext,newtxmath}
\usepackage{orcidlink}

\title[Morphology of flickering magnetised jets]{The impact of flickering variability and magnetisation on the dynamics, stability and morphology of radio-loud AGN jets}
\author[E. L. Elley et al.]{
Emma L. Elley$^{\orcidlink{0009-0002-5349-908X}}$$^{1}$\thanks{E-mail: emma.elley@physics.ox.ac.uk},
James H. Matthews$^{\orcidlink{0000-0002-3493-7737}}$$^{1}$,
Henry Whitehead$^{\orcidlink{0009-0006-0716-0965}}$$^{1,2}$,
Alex J. Cooper$^{\orcidlink{0000-0002-4033-3139}}$$^{1,3,4}$
\\
$^{1}$Astrophysics Subdepartment, Department of Physics, University of Oxford, Keble Road, Oxford, OX13RH, UK \\
$^{2}$Institute of Science and Technology Austria (ISTA), Am Campus 1, Klosterneuburg A-3400, Austria \\
$^{3}$Trottier Space Institute at McGill, 3550 Rue University, Montreal, QC, H3A 2A7, Canada \\ 
$^{4}$Department of Physics, McGill University, 3600 Rue University, Montreal, QC, H3A 2T8, Canada\\
}
\date{Accepted 2026 September 22. Received 2026 September 21; in original form 2026 May 8}

\begin{document}

\maketitle

\begin{abstract}
    The physics governing the morphology of radio-loud AGN jets is not fully understood. We investigate how magnetization, flickering jet power and their interplay affects the morphology of radio galaxies. We present a grid of relativistic magnetohydrodynamic simulations using the PLUTO code covering constant and variable jets with two levels of magnetisation. We find that the constant high magnetisation jets can lead to highly asymmetrical cocoon morphologies, whilst the variable high magnetisation jet can exhibit a broken morphology, caused by a discontinuous jet beam. Our work highlights the importance of magnetisation and variability on the stability and resulting morphology of radio-loud AGN jets, suggesting both are significant factors in addition to jet power or environment. Furthermore, we show that the interaction between magnetisation and variability can lead to the development of localised kink instabilities along the jet beam. Finally, we discuss the effects of hydrodynamic mixing in low magnetisation jets and the role of viewing angle dependence in comparisons between our simulations and observed sources. To facilitate this comparison we present a library of simulated radio images at different times in the simulations and from various viewing angles, which highlight a diverse set of complex morphologies.
\end{abstract}

\section{Introduction}
Observations of radio jets from AGN show a large amount of morphological diversity, with possible causes of differences in age, environment, magnetisation and power. Classically, radio galaxies have been divided into Fanaroff and Riley I and II morphologies~\citep{Fanaroff1974TheLuminosity}, describing centre-brightened and edge-brightened morphologies respectively. Large-scale surveys such as LoTSS~\citep{2022ShimwellLoTSSDR2} have revealed a wealth of jet morphologies, with many sources displaying S- and C-shaped curvatures, multiple hotspots or jets misaligned from the central axis~\citep[e.g.][]{2025HortonComplexDR2}. Moreover, individual nearby sources observed with high spatial resolutions show a wealth of features. For example, Cygnus A~\citep{1989CarilliBroadCygnusA} features a misaligned jet and multiple hotspots, 3C 98 has a misaligned jet together with a one-sided backflow~\citep{1997LeahyStudyCM}, and IC 4296~\citep{2021CondonThreads4296} shows a clear break in the emissivity of the Western jet combined with a helical structure. Understanding the drivers behind how this diversity in structures is formed is critical to understanding the properties of radio galaxies, both within an individual source and across populations. 

Variability in AGN jets is observed over a wide range of timescales. AGN feedback is expected to operate over long timescales, with jet duty cycles estimated at up to approximately 100 Myr~\citep[e.g.][]{2020ShabalaDutyHole,2025QuiciAGNGalaxies}. As we will discuss, within these lifetimes it is possible that the injected jet power may vary in time~\citep[e.g.][]{Gaspari2017RainingModel}. With sufficient X-ray data, average jet power can be estimated from the enthalpy of an inflated cavity~\citep[e.g.][]{2017InesonRepresentativeGalaxies}. Otherwise, average jet power estimates must rely on an extrapolation of the local jet power -- radio luminosity relation. Crucially for studies of jet feedback, if we can understand the formation of the diverse morphologies, this provides an opportunity to study jet launching physics over timescales of $\approx0.1-1$ Myr, and to understand better how to infer average jet powers, for more distant sources where the lobe volume information is not available.

Observations of individual sources give us an insight into a snapshot of the jet's evolution, whereas population-based studies allow us to understand how common various morphologies are. Simulations, then, are a vital tool in advancing our understanding of how various morphologies come to exist, and under which conditions their formation is possible. As an illustration of this, simulations of constant power magnetised jets often display asymmetrical features over large length scales, including various combinations of jet bending, complex hotspot morphologies and one-sided backflows~\citep[e.g.][]{1990HardeeAsymmetricJet,1991CoxThreeSources,2016EnglishNumericalJets, 2019PeruchoLongSimulations,Mukherjee2020SimulatingDynamics,Meenakshi2023PolarizationFields,2024RossiDifferentFlows}.

In RMHD simulations, high magnetisations have been shown to stabilise jets with respect to Kelvin-Helmholtz (KH) instabilities and reduce entrainment, although being cold and fast can also lengthen the growth timescales of KH instabilities~\citep[e.g.][]{2010PeruchoStabilityCollimation}. At the same time, other numerical studies have shown that increasing magnetisation reduces stability to current-driven kink instabilities~\citep{2000ApplCurrentAnalysis}, resulting in the formation of helical patterns in the jet beam~\citep[e.g.][]{Mukherjee2020SimulatingDynamics, 2024RossiDifferentFlows}. \cite{2016TchekhovskoyThreeDichotomy} investigated the potential role of the kink instability in producing the Fanaroff-Riley Dichotomy using 3D MHD simulations, finding it may play an important role, whereas \cite{2024MussoEvolutionJets} studied the stability of more idealised sections of pressure-balanced jets. Furthermore, instability to the kink mode leads to magnetic reconnection which can provide an additional mechanism by which to accelerate electrons in a jet to non-thermal energies~\citep[e.g.][]{2020DavelaarParticleJets,2024MussoEvolutionJets}. \cite{2024RossiDifferentFlows} showed that deceleration due to entrainment is reduced in jets with high magnetisations, enabling the creation of FR II-like morphologies at jet powers of approximately $10^{43}\rm{\,erg\,s}^{-1}$, whereas low magnetisation jets with the same power undergo significant entrainment and deceleration, suggesting the formation of FR I-like morphologies~\citep[see also][]{2008RossiFormationCase}. This picture of low magnetisations in FR I type jets is supported by \cite{2014LaingSystematicGalaxies}, who showed that deceleration of FR I jets occurs before the first recollimation zone; such a result suggests that the deceleration is most likely to be associated with the development of short wavelength instabilities, such as those seen in~\cite{2007PeruchoNumerical31} and previously described in for example \cite{2017MatsumotoLinearFlow, 2018GourgouliatosReconfinementNuclei}.

Whilst magnetisation is known to be a key factor in determining the stability and dynamics of astrophysical jets, variability in injection power is also likely to have significant impacts. Variability in AGN jets is directly observed on short timescales of years, weeks or even hours (e.g.~\citealt{2019ListerJetBlazars,2021ListerMonitoringGalaxies} or for a review, see~\citealt{2025RaiteriVariabilitySpectrum} and references therein). It is expected that AGN jets also vary over very long time scales -- sources such as Hercules A are often explained with restarting jets~\citep[e.g.][]{2022TimmermanOriginObservations}, and AGN jets are expected to be influenced by (and to influence) merger activity~\citep[e.g.][]{2024TalbotSimulationsMergers}. Further to this, theories such as chaotic cold accretion~\citep[CCA,][]{king2006,Gaspari2013ChaoticHoles} suggest that the accretion rate should vary on intermediate timescales of fractions of a Myr with a lognormal distribution of powers and obeying a pink noise spectrum~\citep{,Gaspari2017RainingModel}. In this work we focus on this intermediate timescale, pink noise variability ($\approx0.1-1$ Myr), which we describe as `flickering'. Much shorter timescales are below the time resolution of our simulations, and are also expected to be smoothed out by internal shocks~\citep{Malzac2014TheShocks} and other dissipation mechanisms on length scales below the resolution of our simulations~\citep[see section 4.2 of][]{2026ElleySimulatingBrightening}.

In \cite{2026ElleySimulatingBrightening} we investigated flickering jets using relativistic hydrodynamic (RHD) simulations, finding that moderate changes in Lorentz factor on timescales of a fraction of a Myr could create large short-lived increases in synchrotron luminosity. However, variability is of course not limited to flickering -- in the context of restarting jets,  \cite{2017DuranSimulationsShocks} found that stability to the kink mode can be greatly affected by previous epochs of jet activity. \cite{2026RatnaparkhiNumericalDynamics} have studied the differences in the propagation and dynamics of FRII jets when launched into a volume created by a previous episode of jet activity, whereas~\cite{2026JlassiSimulatingCluster} have focused on modelling variable power FRI jets in a cool-core cluster. Furthermore, variability can also refer to changes to variables other than a jet's power -- \cite{2023ChenNumericalEvolution} study the evolution of magnetised jets, including a jitter in the initial launch angle of the jets. The magnetised jets with jitter display complex morphologies with large scale twisting magnetic field structures directly imprinted by the varying jet direction.

To summarise, previous work has suggested that both variability and magnetisation can have significant effects on the morphologies and stability of AGN jets, and that variability in the morphologies and luminosities of individual jets may contribute strongly to the observed diversity of sources. In particular, some degree of flickering variability is an expected consequence of accretion theories and can lead to variability in jet morphology in non-magnetised jets. In order to exploit the potential information encoded within jet morphologies, it is necessary to develop a detailed understanding of the possible ways in which jets with different magnetisations and variability properties can evolve. As of yet, no systematic investigation has been carried out on the interplay between variability and magnetisation and their joint impacts on the jet-lobe system. In this work, we study this combination of flickering variability with dynamically important magnetic fields and assess consequences for the expected morphologies of powerful radio sources. We use a grid of simulations with and without variability, and with low and high magnetisations to separate the effects of each property. It should be noted that we refer to the $\sigma_{B,\rm{peak}}\left(\bar{Q_j}\right)=0.01$ jet simulations as `high' magnetisation in comparison to our low magnetisation jets, but both the low and high magnetisation jets correspond to small fractions of the energy flux being contained in the magnetic fields. 

We describe our simulations in Section~\ref{sec:methods}, paying particular attention to the model of a magnetised jet used, together with the specific properties of the variability. In Section~\ref{sec:results}, we discuss the results of the simulations, giving an overview of the results from the high magnetisation, variable power jet in Section~\ref{sec:results_high_variable}, before focusing on large-scale asymmetries in Section~\ref{sec:results_asymmetries}, and on the formation of discontinuities in the jet beam in Section~\ref{sec:discontinuities_results}. In Section~\ref{sec:discussion_stability}, we discuss the stability of our injected jet model to $m=1$ kink instabilities and how this governs the formation of helical structures and localised kink structures in the jets. Section~\ref{sec:predicted_images} presents simulated radio images made using the simulations, together with a discussion of viewing angle effects, light travel time effects and comparisons with observations of specific sources. We conclude in Section~\ref{sec:conclusions}.
\section{Methods}
\label{sec:methods}
\subsection{Simulation setup}
We run 3D RMHD simulations of jets with both constant and varying power using the PLUTO code~\cite{Mignone2007PLUTO:Astrophysics}.  The domain of each simulation is $60 \rm{\,kpc}\times60 \rm{\,kpc}\times60 \rm{\,kpc}$ and consists of $1500\times1500\times1500$ cells, giving a resolution of 40 pc per cell in each direction. We inject a jet along the $z$ axis at $z=0$.
\begin{table}
    \centering
    \begin{tabular}{c|c|c}
    \hline
         Power & Magnetisation & $\sigma_{B,\rm{peak}}\left(\bar{Q_j}\right)$\\ \hline
         Constant & Low & 0.001 \\
         Constant & High & 0.01 \\
         Variable & Low & 0.001 \\
         Variable & High & 0.01 \\ \hline
    \end{tabular}
    \caption{Overview of parameters for the four simulation runs. $\sigma_{B,\rm{peak}}\left(\bar{Q_j}\right)$ refers to the magnetisation at $r=a=0.2$ kpc for the mean jet power.}
    \label{tab:sim_params}
\end{table}

The variable injection power jets have powers which obey a lognormal probability density function
\begin{equation}
    p(Q_j) = \frac{1}{\sqrt{2\pi}\times0.33 Q_j}\exp\left[-\frac{(\ln Q_j - \ln Q_{j,0})^2}{2\times0.33^2}\right]\, ,
    \label{eqn:jet_power_pdf}
\end{equation}
where $Q_j$ is the jet power at any given time, $Q_{j,0}$ is the median jet power and the standard deviation of the natural logarithm of the jet powers is 0.33. The jet power time series has a pink noise power spectrum (power spectral density $S(f)\propto f^{-1}$). We generate the time series for the jet power using DELightcurveSimulation~\citep{Connolly2016DELightcurveSimulation:Code} which is an implementation of the Emmanoulopoulos light curve simulation algorithm~\citep{Emmanoulopoulos2013GeneratingUpdated}. The grid of simulations consists of 4 runs with varying combinations of magnetisation and variability as shown in Table~\ref{tab:sim_params}. The variable simulations use the same random noise seed as Seed 6 as in~\cite{2026ElleySimulatingBrightening}.

We use linear reconstruction, second-order Runge-Kutta time-stepping and the Taub-Mathews Equation of State~\citep{1971MathewsHydromagneticGas}. The divergence-free condition is maintained through divergence cleaning. Away from strong shocks, we use the Harten–Lax–van Leer Riemann (HLLD) solver of ~\cite{2009MignoneFiveMagnetohydrodynamics} with the monotonized central difference limiter. We use the multidimensional (MULTID) shock identification and flattening strategy, where for strong shocks above a threshold of $\epsilon_{\rm sh} \equiv \left|\frac{\Delta P}{P}\right|> 0.5$ (where $P$ is the gas pressure), the (diffusive) minmod limiter is used for reconstruction and the Harten–Lax–van Leer (HLL) Riemann solver is used. The MULTID approach therefore forces the simulation to be more dissipative around shocks, but is required for numerical stability.

Initially the ambient density is set to a King profile,
\begin{equation}
    \rho(r) = \rho_0\left[1+\left(\frac{r}{r_c}\right)^2\right]^{-\frac{3\beta}{2}}
\end{equation}
with a core radius, $r_c$, of 50 kpc, and with $\beta = 0.5$ and $\rho_0=6.0\times10^{-27}$ g cm$^{-3}$. $\rho_0$ is equivalent to a proton number density of $n_p\approx3.5\times10^{-3}\,\rm{cm}^{-3}$ and our values for $r_c$, $\beta$ and $n_p$ are broadly consistent with typical values for environments of radio-loud AGN~\citep[e.g.][]{2015InesonLinkGalaxies}. In particular, our values of $\beta$ and $r_c$ are close to those estimated for PKS 2356--61 ($\beta=0.42$ and $r_c=56.34$) and 3C 388 ($\beta = 0.52$ and $r_c = 40.39$), which have an estimated cluster radius, $R_{500}$, of $465$ kpc and $866$ kpc respectively, where $R_{500}$ is the radius within which the mean density is $500$ times the critical density of the Universe. These correspond to cluster/group masses in the range $M_{500}\approx4.8\times10^{46}-3.1\times10^{47}\rm{\,g}\approx2\times10^{13-14}\rm{\,M_\odot}$. Small perturbations are added to this density profile to remove the exact symmetry of the simulation. The pressure of the ambient medium is set to a low value ($p= 3.2\times10^{-14}\rm{\,dyn\,cm}^{-2}$) equivalent to a temperature of $\approx4eV$. Through 2D axisymmetric RHD tests, we have verified that the propagation speed and cocoon shape are not significantly impacted when the ambient pressure is increased by a factor of 1000, which brings the temperature in the ambient medium in line with estimates of the ICM temperature in environments hosting FRII radio galaxies~\citep[e.g.][]{2017InesonRepresentativeGalaxies} ($\approx1.19$ keV for PKS 2356--61 and $\approx3.52$ keV for 3C 388). For numerical reasons, we set the magnetic field in the background to a small but non-zero, dynamically negligible constant value, in a direction perpendicular to the jet axis (See Appendix~\ref{sec:cartesian_appendix}).  A more realistic background field is outwith the scope of this work -- we focus on the field structures developing within the jet. The bow shock ensures that the low magnitude magnetic field structure in the ambient medium has little effect on the internal magnetic field structures.

We use outflow boundary conditions on all sides, except the side through which the jet enters the simulation; this boundary is set to be reflective everywhere but the jet nozzle. We inject a jet with a density of $\rho_j=10^{-4}\rho_0=6.0\times10^{-31}$ g cm$^{-3}$ through a nozzle of radius $r_j=0.4$ kpc. We discuss the remaining boundary conditions for the jet in Section~\ref{sec:methods_jet_model}. 

\subsection{Jet model}
\label{sec:methods_jet_model}
We simulate an underdense jet with a toroidal magnetic field. The jet is initially in radial equilibrium, which requires the specification of a magnetic field (Section~\ref{sec:field_structure}) and pressure profile (Section~\ref{sec:methods_pressure_profile}) within the jet. In order to create a jet with the required power at a given level of magnetisation (Section~\ref{sec:methods_magnetisation}), we vary the Lorentz factor as described in  Section~\ref{sec:Lorentz_factor_from_power}. We use the convention that primed quantities are as measured in the rest frame of the fluid throughout.
\subsubsection{Magnetic field structure}
\label{sec:field_structure}
We choose a toroidal field structure adhering to some general constraints, namely that the field must have zero divergence ($B_r=0$) and not result in a singularity at the centre of the jet ($|B(r=0)|=0$). Furthermore, we stipulate that the field strength must fall off to a small value at the outer radius of the jet. This avoids a large step change at the outer radius, improving numerical stability and is a reasonable physical expectation. The simplest non-trivial structure obeying these constraints is a field that reaches a maximum strength at some critical cylindrical radius $a<r_j$ within the jet, before decreasing in strength with further increasing radius. This is qualitatively similar to the toroidal component of the helical fields used in for example~\cite{2009MizunoThreeColumn, 2021BodoKinkBLLacs} and the toroidal field used in, for example,~\cite{Mukherjee2020SimulatingDynamics}. We make the choice that $a=r_j/2$ and describe fields that reach a maximum strength at this point. Whilst the exact profiles and value of $a$ are specific choices, we note that other reasonable choices are unlikely to greatly alter the overall qualities of our results. To motivate this expectation, we analyse the stability of this outer region of the jet in Appendix~\ref{sec:appendix_stability_outer}, showing that for the same small displacement from the equilibrium, the resulting forces on the outer region are much smaller than those experienced by the central region (an analysis of which can be found in Section~\ref{sec:discussion_stability_inner}).

Within the radius $r=a$, we inject a toroidal field with
\begin{equation}
    B_\phi=\frac{r}{a}B_{\phi, \rm{\,max}},
\label{eqn:Bfield_within_cylindrical}
\end{equation}
 and $B_r=B_z=0$. Outside of the radius $r=a$ we use a field with
 \begin{equation}
 B_\phi=\frac{a}{r}B_{\phi, \rm{\,max}},
 \label{eqn:Bfield_outwith_cylindrical}
 \end{equation}
 and again $B_r=B_z=0$. The two field profiles result in an equal strength at $r=a$. For $a<r<r_j$, the field given above has a decreasing strength with radius, to a low value at $r_j$. The Cartesian versions of these fields and the smoothing function used at the outer radius of the injected jet, $r_j$, are described in Appendix~\ref{sec:cartesian_appendix}.

\subsubsection{Equilibria}
\label{sec:methods_equilibria}
The field structure discussed in Section~\ref{sec:field_structure} results in a magnetic pressure inwards on the jet, so to create an equilibrium this force must be balanced by either rotation of the jet or a pressure gradient, as shown below. As we focus on simulating jets on scales of kpc to tens of kpc in this work, which are scales at which the jet material is distant from the jet launching region and at which the jet is expected to have already transformed most of its Poynting flux to thermal and kinetic energy, we focus on equilibria where the remaining magnetic tension is balanced purely by pressure gradients and not by any rotation of the jet~\citep[e.g.][]{2024MussoEvolutionJets}.

For a steady-state jet assuming ideal magnetohydrodynamics (MHD) with negligible resistivity, the condition for an equilibrium state is given by 
\begin{equation}
    \frac{d}{dr}p(r)+\frac{1}{r^2}\frac{d}{dr}\left[r^2\frac{B_\phi'^2(r)}{8\pi}\right] -\rho'\frac{v_\phi^2}{r}=0,
    \label{eqn:equilibrium_condition}
\end{equation}
where we have used the assumption that $p$ and $\vec{B}'$ vary only in the $r$ direction. The first and second terms represent the gradients of the thermodynamic and magnetic pressures respectively and the third the centripetal force needed to maintain a circular rotation (if the toroidal velocity were non-zero). In this work we consider non-rotating jets, such that $v_\phi^2=0$. Furthermore, in ideal MHD, the fluid is perfectly conductive such that
\begin{equation}
    \vec{E'}=-\vec{v'}\times\vec{B'}
\end{equation}
For a non-rotating jet where the velocity does not vary with radius, there is a co-moving frame where $v'=0$ throughout a slice of the jet, such that the electric field in the co-moving frame is zero. To convert between the rest frame and comoving frame toroidal fields, we note that the toroidal component of the magnetic field in the observer frame is
\begin{equation}
    B_\phi=\Gamma\left(B_\phi'-\frac{v_j}{c^2}E_r'\right)=\Gamma B_\phi',
\end{equation}
assuming $v_\phi=0$.
\subsubsection{Pressure profile}
\label{sec:methods_pressure_profile}
In the observer frame, Equation~\ref{eqn:equilibrium_condition} can be written as
\begin{equation}
    \frac{dp}{dr}+\frac{1}{8\pi r^2\Gamma^2}\frac{d}{dr}\left(r^2B_\phi^2(r)\right)=0
\label{eqn:specific_equilibrium}
\end{equation}
For $r>a$, this necessitates a constant pressure with changing radius. For $r<a$, we obtain a similar result to~\cite{Mignone2013ModellingNebula} for the required thermal pressure at a given time:
\begin{equation}
    p_j=-\frac{r^2B_{\phi, \rm{\,max}}^2}{4\pi\Gamma^2a^2}+\rm{const} = -\frac{B_\phi'^2}{4\pi}+\rm{const}
\end{equation}
We follow~\cite{Mignone2013ModellingNebula} in defining the constant such that the jet pressure matches the ambient pressure at a radius $r = a=r_j/2$.
\begin{equation}
    p_j=p_a+\frac{B_{\phi, \rm{\,max}}^2}{\Gamma^2}\left(1-\frac{r^2}{a^2}\right)=p_a+B_{\phi, \rm{\,max}}'^2\left(1-\frac{r^2}{a^2}\right)
    \label{eqn:pressure_profile}
\end{equation}
\subsubsection{Magnetisation}
\label{sec:methods_magnetisation}
We adopt the usual definition \citep[e.g.][]{Mukherjee2020SimulatingDynamics} of the jet magnetisation as the ratio between the Poynting flux, $\vec{S_j}= \vec{B_j}\times(\vec{v_j}\times \vec{B_j})/4\pi$, and the enthalpy density flux, $\vec{F_j}=(\Gamma^2\rho_jh_j-\Gamma\rho_jc^2)\vec{v_j}$,
\begin{equation}
    \sigma_B=\frac{|\vec{S_j}\cdot\hat{z}|}{|\vec{F_j}\cdot\hat{z}|}=\frac{|(\vec{B_j}\times(\vec{v_j}\times \vec{B_j}))\cdot\hat{z}|}{4\pi(\Gamma^2\rho_jh_j-\Gamma\rho_jc^2)(|\vec{v_j}\cdot\hat{z}|)},
    \label{eqn:magnetisation}
\end{equation}
where $\rho_jh_j$ is the relativistic enthalpy density of the jet, given by 
\begin{equation}
    \rho_jh_j=\frac{5}{2}p_j+\sqrt{\frac{9}{4}p_j^2+(\rho_jc^2)^2},
\end{equation}
according to the Taub-Mathews Equation of State~\citep{1971MathewsHydromagneticGas}. We neglect the rest mass contribution to the enthalpy flux. The peak value of the magnetisation occurs at $r=a$ and is given by
\begin{equation}
    \sigma_{B,\rm{peak}}=\frac{B_{\phi, \rm{\,max}}^2}{4\pi\Gamma^2\left(\frac{5}{2}p_a+\sqrt{\frac{9}{4}p_a^2+(\rho_jc^2)^2}\right)}.
    \label{eqn:peak_magnetisation}
\end{equation}
In the variable injection power simulations we hold the injected lab frame magnetic field strength constant, leading to a varying magnetisation.

\begin{table}
    \centering
    \begin{tabular}{c|c|c}
    \hline
         $\sigma_{B,\rm{peak}}$ & 0.001 & 0.01\\ \hline
         $\Gamma_j$ & 4.17 & 4.10 \\
         $v_j$ & 0.971c & 0.970c \\
         $B_{\phi, \rm{\,max}}\,(\rm{\mu G})$ & 11 & 34 \\
         Integrated Enthalpy Flux ($\rm{erg\,s}^{-1}$) & $1.0\times10^{45}$ & $1.0\times10^{45}$\\
         Integrated Internal Energy Flux ($\rm{erg\,s}^{-1}$) & $1.0\times10^{45}$ & $9.8\times10^{44}$\\
         Kinetic Power ($\rm{erg\,s}^{-1}$) & $2.4\times10^{44}$ & $2.3\times10^{44}$\\
         Magnetic Power ($\rm{erg\,s}^{-1}$) & $4.7\times10^{41}$& $4.5\times10^{42}$\\\hline
    \end{tabular}
    \caption{Properties of the jet at injection for the constant injected power cases, with fluxes integrated across the jet radius. All fluxes are integrated across the area of the jet. A very small fraction of the jets' power is in the form of Poynting flux.}
    \label{tab:jet_properties}
\end{table}

\subsubsection{Lorentz factor from the power of the jet}
\label{sec:Lorentz_factor_from_power}
The jet power is given by the integral of the total enthalpy flux over the area of the jet nozzle, including the magnetic field contribution, and is given by
\begin{multline}
Q_j=2\pi\int^{a}_0\left(\left|\vec{S}_{j,\rm{in}}\cdot\hat{z}\right|+\left|\vec{F}_{j,\rm{in}}\cdot\hat{z}\right|\right)rdr\\+2\pi\int^{r_j}_a\left(\left|\vec{S}_{j,\rm{out}}\cdot\hat{z}\right|+\left|\vec{F}_{j,\rm{out}}\cdot\hat{z}\right|\right)rdr, 
\end{multline}
where the `in' and `out' subscripts refer to the quantities within $0\leq r\leq a$ and $a<r\leq r_j$ respectively. Substituting in the expressions for the Poynting and enthalpy fluxes (see Equation~\ref{eqn:magnetisation}), and again neglecting the contribution from the rest mass energy to the enthalpy flux, gives
\begin{multline}
    Q_j=2\pi\int^{a}_0\frac{B_{\phi, \rm{\,max}}^2r^3}{4\pi a^2}v_j+\Gamma^2\left(\frac{5}{2}p_j+\sqrt{\frac{9}{4}p_j^2+(\rho_jc^2)^2}\right)rv_jdr\\+2\pi\int^{r_j}_a\frac{B_{\phi, \rm{\,max}}^2a^2}{4\pi r}v_j+\Gamma^2\left(\frac{5}{2}p_a+\sqrt{\frac{9}{4}p_a^2+(\rho_jc^2)^2}\right)rv_jdr.
    \label{eqn:jet_power_integral}
\end{multline}

\begin{figure}
    \centering
    \includegraphics[width=\linewidth]{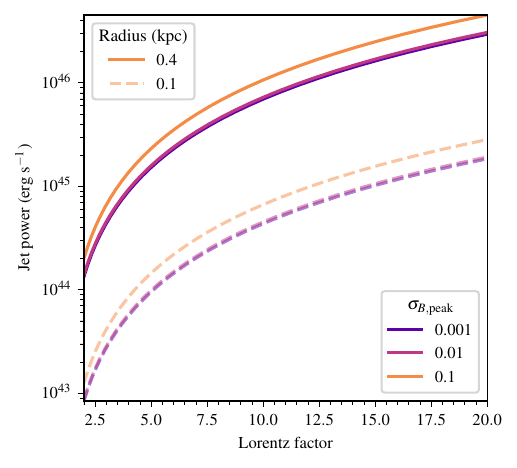}
    \caption{Jet power as a function of Lorentz factor for various values of the peak magnetisation ($\sigma_{B,\rm{peak}} $) and for jets with radius $r_j = 0.4$ kpc (solid lines) and $r_j = 0.1$ kpc (dashed lines). The jets we simulate in this work have a radius of $0.4$ kpc. The curves for $\sigma_{B,\rm{peak}}= 0.01$ and $\sigma_{B,\rm{peak}}= 0.001$ have approximately equal values at all points.}
    \label{fig:lorentz_to_power}
\end{figure}

\begin{figure}
    \centering
    \includegraphics[width=\linewidth]{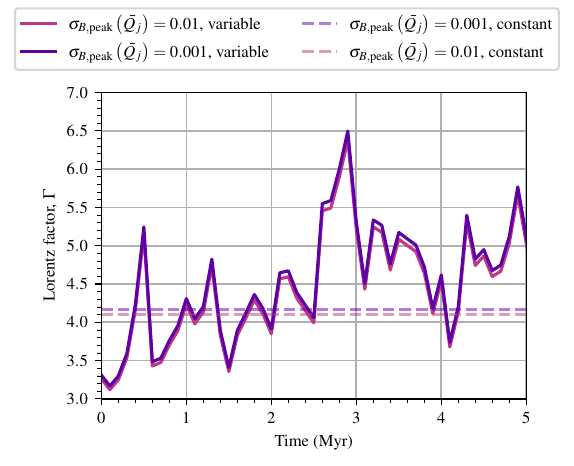}
    \caption{Lorentz factor as a function of time for the four simulations, colour-coded by our magnetisation parameter,  $\sigma_{B,\rm{peak}}$. Solid lines show the Lorentz factors used for the variable jets, whilst dashed, fainter lines show the Lorentz factors for the steady jets. As can be seen from Figure~\ref{fig:lorentz_to_power}, for small magnetisations, the difference in Lorentz factor required for a given power is very small, with the more highly magnetised jet requiring a slightly lower Lorentz factor to have the same power.}
    \label{fig:Lorentz_factors_timeseries}
\end{figure}

To parametrise our jets, we specify the mean jet power, $\bar{Q}_j$, together with the peak magnetisation at this mean jet power, which we denote $\sigma_{B,\rm{peak}}\left(\bar{Q}_j\right)$. Equation~\ref{eqn:peak_magnetisation} specifies the comoving peak toroidal magnetic field strength, which is independent of Lorentz factor. This can then be substituted into Equation~\ref{eqn:jet_power_integral} with the appropriate Lorentz transformation to find the power of a jet with this comoving magnetic field strength. To invert Equation~\ref{eqn:jet_power_integral}, we tabulate the power for a range of  Lorentz factors using Equation~\ref{eqn:jet_power_integral} and use this data as a look-up table to find the Lorentz factor as a function of power. Figure~\ref{fig:lorentz_to_power} shows the jet power as a function of Lorentz factor for various levels of magnetisation. 
It is notable that for the magnetisations we use ($\sigma_{B,\rm{peak}}\left(\bar{Q_j}\right)=0.001,0.01$, as shown in Table~\ref{tab:sim_params}), the jet power depends only very weakly on the magnetisation. This justifies our choice to use the magnetisation at the mean jet power to calculate the Lorentz factor for all Lorentz factors. Figure~\ref{fig:Lorentz_factors_timeseries} shows the Lorentz factors calculated using the method. In Table~\ref{tab:jet_properties}, we list the fluxes and other key jet properties for the constant injected power low and high magnetisation simulations.

An alternative possible choice would be to hold the magnetisation constant and instead vary the injected toroidal magnetic field strength in the lab frame. In this case, the comoving injected toroidal field would remain constant, and therefore the injected pressure would also be held constant in time.

\subsection{Production of rendered emissivity images}
\label{sec:rendering}
As part of our work, we also create predictions of observed synchrotron images with an approximate treatment of the synchrotron emissivity of the plasma. We do this from various angles to the line of sight to illustrate the effects of Doppler boosting on the appearance of the jets, using the open-source ray tracing code cuDART~\citep{2026WhiteheadcuDARTSources}. To produce the emissivity images from different angles, we first transform the magnetic fields to the co-moving frame for each cell, then calculate rest frame emissivity estimates for each cell. We use the prescription that pseudo-emissivity (in the rest frame) is given by~\citep[e.g.][]{1985RybickiRadiativeAstrophysics}
\begin{equation}
    j'(\omega_0')\propto B'^{1.6}u_B'\mathcal{C}\propto B'^{3.6}\mathcal{C},
    \label{eqn:pseudo_emiss_prop}
\end{equation}
where $\omega_0'$ is a reference emitting frequency, $u_B'$ is the magnetic energy density and $\mathcal{C}$ is the jet tracer which tracks the fraction of material in a cell which entered the simulation through the jet nozzle. Our approach does not account for radiative losses. We do not specify an observing frequency that our simulated radio images are made at, but assume that it is low enough to be well-described by power law emission with a single slope. We assume that the electron energy density spectrum is given by a power law with an index of $p = 2.2$, and that the resulting synchrotron spectrum has an index $\alpha = -0.6$, such that $F'(\nu)\propto\nu^{-0.6}$. We describe the processes of calculating the co-moving magnetic field strengths and converting them into an observed simulated radio image at a single observing frequency in further detail in Appendix~\ref{sec:rendering_appendix}. We calculate the emergent specific intensity using an in-house ray tracing code to perform line-of-sight integration of the pseudo-emissivity in the observer frame, Doppler boosting the emissivity on a per-cell basis according to the local fluid velocity with respect to the line of sight (see Appendix~\ref{sec:rendering_appendix}).  

We simulate one jet (or one side of a radio galaxy) in each case, whereas almost all observed radio galaxies are seen to be double-sided, with one-sided jets commonly explained by strong Doppler de-boosting of the receding jet~\citep[e.g.][]{2025HortonComplexDR2}. To produce the double-sided jets shown in our emissivity maps we mirror the jets along the jet axis and then, for the mirrored jet, we swap the two axes perpendicular to the jet axis. We carry out this axis-swap to reduce the symmetry and allow a wider variety of views of the simulated jet overall. We do not account for light travel time effects as this would require simulation data to be saved much more often during the simulation, which is not feasible due to storage and processing limitations. We discuss light travel time effects further in Section~\ref{sec:viewing_angle_effects}. All scales shown in kpc on images throughout this paper are distances as projected onto the plane of the sky.

We release a set of rendered images together with this paper. These cover the following views to the simulated objects:
\begin{enumerate}
    \item The jet axis remains perpendicular to the line of sight ($\theta = 90^\circ$). The simulation is rotated around the jet axis in $\Delta\phi=3$ degree increments.
    \item The angle between the jet axis and the line of sight, $\theta$, is varied in 3 degree increments for a single constant azimuthal angle, $\phi$.
\end{enumerate}
We discuss some of the effects of viewing the same jet from different angles in Section~\ref{sec:viewing_angle_effects} and videos of the highly magnetised jets can be viewed here: \href{https://youtu.be/JSFDY7h8VXY}{constant} and \href{https://youtu.be/k22vz1wRQW0}{variable}.
\section{Results}
\label{sec:results}
We discuss the broad evolution of the high magnetisation jet with variable injection power in Section~\ref{sec:results_high_variable}. We then move to comparing the variable high magnetisation jet to its low magnetisation and constant counterparts; this allows us to identify which features are characteristic of high magnetisation, which are characteristic of variability and which are a consequence of the interplay between the two. We summarise the key features of all the simulations in Table~\ref{tab:sim_results}; large-scale asymmetries are discussed in Section~\ref{sec:results_asymmetries}, while discontinuous or `broken' morphologies are explored in Section~\ref{sec:discontinuities_results}.
\subsection{Evolution of the high magnetisation variable jet}
\label{sec:results_high_variable}
\begin{figure*}
    \centering
    \includegraphics[width=\linewidth]{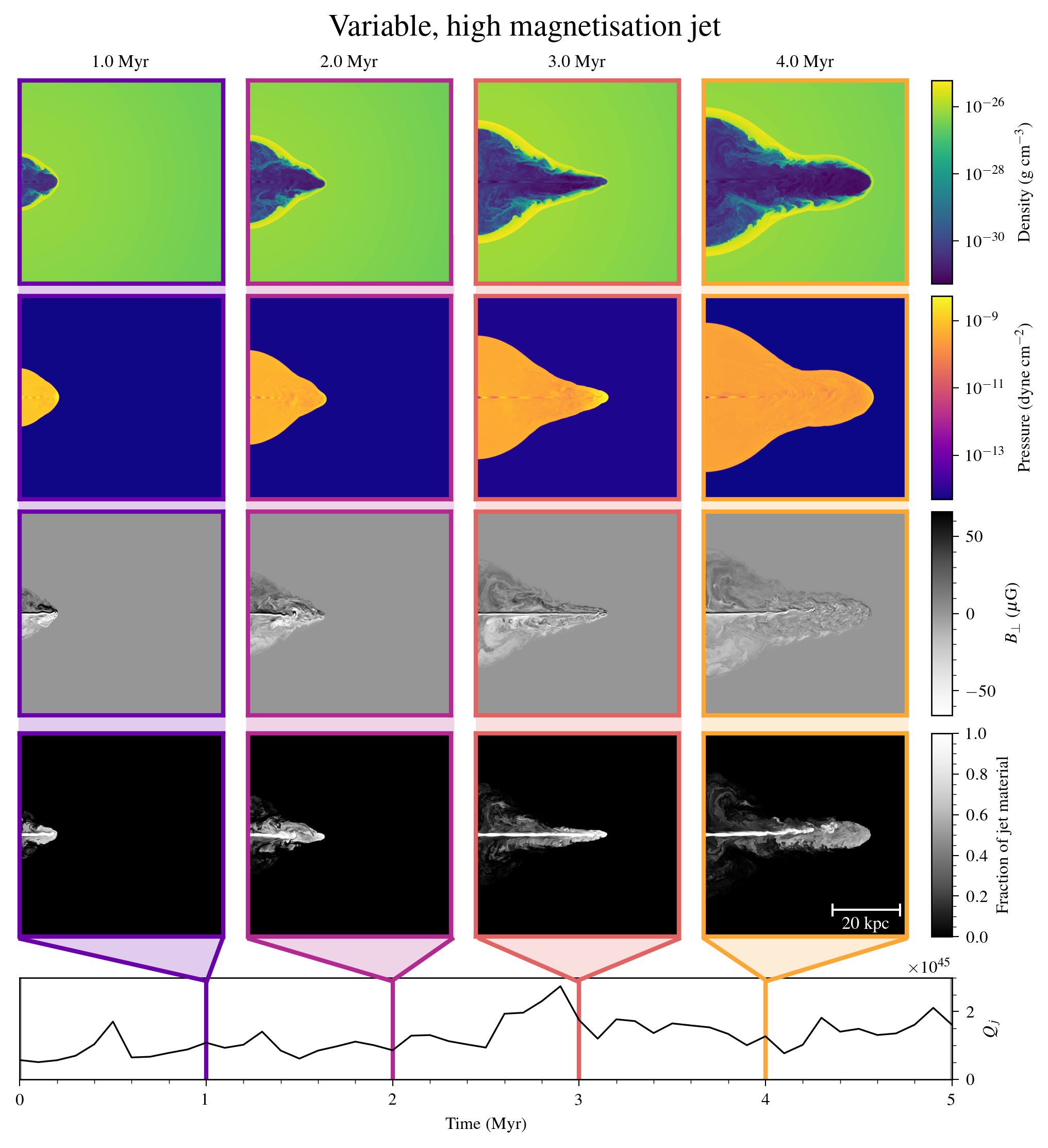}
    \caption{Central slices through the variable injection power, high magnetisation simulation showing the density, pressure, the component of the magnetic field perpendicular to the plane ($B_\perp$), which is equivalent to the toroidal component at the jet base and the fraction of jet material (jet tracer). The full size of the slice is 60 kpc by 60 kpc. The cocoon shape of the variable jet shows a strong dependence on the recent power of the jet and the jet can be seen to move significantly away from the central axis around 4 Myr.}
    \label{fig:summary_seed6_010}
\end{figure*}

We begin by discussing the overall behaviour of the variable injection power, high magnetisation jet simulation, which we illustrate in Figure~\ref{fig:summary_seed6_010}. We show central slices through the full three dimensional domain of the simulation, illustrating the density, pressure, component of the magnetic field perpendicular to the slice and the jet tracer, $\mathcal{C}$. The figure shows the underdense, magnetised jet propagating through the majority of the 60 kpc domain in around 4 Myr, with significant variation in the cocoon shape across this time.

At 2 Myr the jet shows some large ($\approx5$ kpc) scale asymmetry around the jet head. This is most clearly seen in the density slice. Whilst the tracer shows the jet beam to be slightly bent away from the central axis at this time, it also shows that the jet beam appears disconnected from the head of the jet at this time, and this disconnected region coincides spatially with a more complicated structure in the magnetic field, which separates the well-defined toroidal field structure that is injected at the base from a more turbulent structure around the jet head region. In the region just before the complex break the magnetic field has small bends reminiscent of structures formed by the kink instability in previous works~\citep[e.g.][]{2021BodoKinkBLLacs,Meenakshi2023PolarizationFields}.

At 3 Myr the jet has experienced a recent extended period ($\approx 0.5$ Myr) of relatively high power and its shape towards the jet head is narrow and pointed. The formation of this sharp, conical region is dominated by the increased ram pressure, concentrated in the area around the jet head -- a consequence of the additional momentum flux of the jet at this time. This behaviour was discussed in detail in the context of non-magnetised jets in~\cite{Whitehead2023StudyingJets}. In this snapshot, the magnetic field structure and the tracer image both show that the jet reaches the end of the simulation domain (largely) unimpeded by instabilities. The thermal pressure at the jet head is significantly increased with respect to its value at both 2.0 and 4.0 Myr, reminiscent of the behaviour of the non-magnetised jets discussed in~\cite{2026ElleySimulatingBrightening}, where changes in power lead to short-lived periods of dramatically brighter hotspots, as a result of the travelling shocks formed by the variability interacting with the pre-existing shock structure at the jet head. The increased pressure in the highly magnetised jet shown in Figure~\ref{fig:summary_seed6_010} suggests that hotspot brightening as a result of changes in jet power may continue to play an important role in jets with dynamically significant magnetic field strengths.

At 4 Myr, following a period of comparatively low jet power, the cocoon shape has returned to a blunter morphology with a higher sphericity. The primary mechanism causing the development of the higher sphericity is the lateral expansion of the cocoon due to its thermal pressure, which plays a more important role than the focused ram pressure in phases of relatively low power. The thermal pressure is much more uniform across the cocoon. Furthermore, the effects of hydrodynamic instabilities are much more evident in the mixing of the jet material both in places along the jet beam and in the jet head area. The jet tracer shows that the jet has moved away from the central axis towards the top of the figure. From 2D slices, it is not possible to tell whether the jet is truly discontinuous, or whether it has simply been bent out of the plane of the slice. Therefore, in Section~\ref{sec:discontinuities_results}, we discuss the discontinuous behaviour of the highly magnetised variable jet simulation in more detail, using volume renders of the jet tracer instead of the slice shown here (see Figure~\ref{fig:broken_jets}). The volume renders allow us to show that the jet is in fact discontinuous at this point.

\begin{table*}
    \centering
    \begin{tabular}{c|c|c|c}
    \hline
         Description & Moves off axis? & Variable morphology? & Backflows\\ \hline
         Constant power, low magnetisation & No & constant & symmetric\\
         Constant power, high magnetisation & Yes, for extended periods & variable, prominent lobe asymmetries & often one-sided\\
         Variable power, low magnetisation & No & variable, travelling shocks & symmetric, vary in strength\\
         Variable power, high magnetisation & Yes, for short periods & variable, broken morphology & often one-sided, clumped\\ \hline
    \end{tabular}
    \caption{Comparison of the key features of the 4 simulations. See Table~\ref{tab:sim_params} for parameters used in each run. Section~\ref{sec:discontinuities_results} focuses on the broken morphologies. Movement away from the central axis is discussed in Section~\ref{sec:results_asymmetries}.}
    \label{tab:sim_results}
\end{table*}
\subsection{Large-scale asymmetries in high magnetisation jets}
\label{sec:results_asymmetries}
Our two high magnetisation ($\sigma_{B,\rm{peak}}=0.01$) jet simulations exhibit departures from axisymmetry on large length scales, $\gtrsim10$ kpc, in agreement with previous work focusing on simulations of constant power jets with toroidal and helical fields~\citep[e.g.][]{2010MignoneHighJets,Mukherjee2020SimulatingDynamics,2023ChenNumericalEvolution,Meenakshi2023PolarizationFields, 2024RossiDifferentFlows}. The asymmetrical behaviour we see can be grouped into two categories: (i) asymmetrical pointing of the jet beam towards one side of the lobe (misaligned jets) and (ii) asymmetries in the cocoon shape, which when they occur happen as a consequence of behaviour (i) persisting for extended periods of time. Large-scale asymmetries are not generally produced in the low magnetisation ($\sigma_{B,\rm{peak}}=0.001$) jets, which remain approximately symmetrical on these $\gtrsim10$ kpc scales (e.g. Figure~\ref{fig:time_comparison_001}, and in agreement with the results of~\citealt{2023ChenNumericalEvolution}).

\begin{figure}
    \centering
    \includegraphics[width=\linewidth]{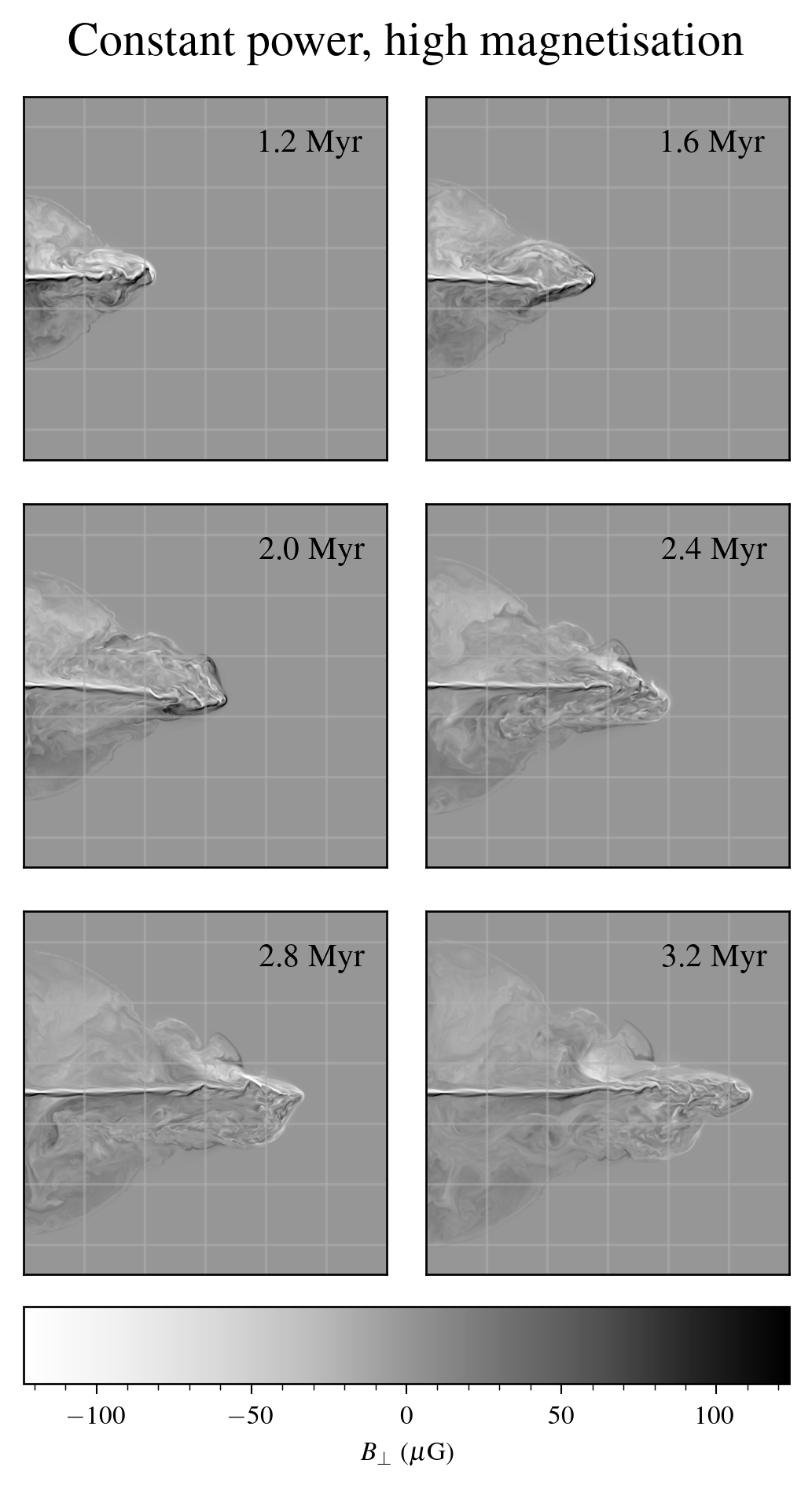}
    \caption{Central slices through the constant power, high magnetisation jet simulation, showing the component of the magnetic field perpendicular to the slice at 6 timestamps. Near to the jet base, this quantity is equivalent to the toroidal field strength into and out of the page. This figure illustrates the development of the asymmetrical cocoon structure at the jet head. The box size shown is 40 kpc in each direction. A video of this can be viewed here: \href{https://youtube.com/shorts/NaL5jEjbO4I}{Bx1\_constant\_010}}
    \label{fig:Bx1_constant_010}
\end{figure}

Figure~\ref{fig:Bx1_constant_010} shows a central slice through the toroidal magnetic field component of the high magnetisation, constant power jet, at six time stamps between 1.2 and 3.2 Myr.
At around 2 Myr the jet moves away significantly from the central axis. The ram pressure of the jet is concentrated on the area immediately around the end of the jet beam and, over a period of $\sim1$ Myr, preferentially excavates a significant volume of cocoon away from the central axis. The jet then returns to the central axis of the simulation. At 2.4 Myr, the backflow can be seen to travel preferentially into the previously excavated side creating a one-sided backflow (see the video of the magnetic field included in the supplementary material: Bx1\_constant\_010.mp4). Between 2.8 and 3.2 Myr the magnetic field strength maps show evidence of the majority of the ram pressure being imparted along the jet axis, leading to a relatively thin cocoon shape at the jet head, with the previously evacuated region still visible towards the bottom of the images. We have verified the alignment of the extension of the lobe with the central axis of the simulation by examination of the corresponding perpendicular central slices through the simulation.

\begin{figure}
    \centering
    \includegraphics[width=\linewidth]{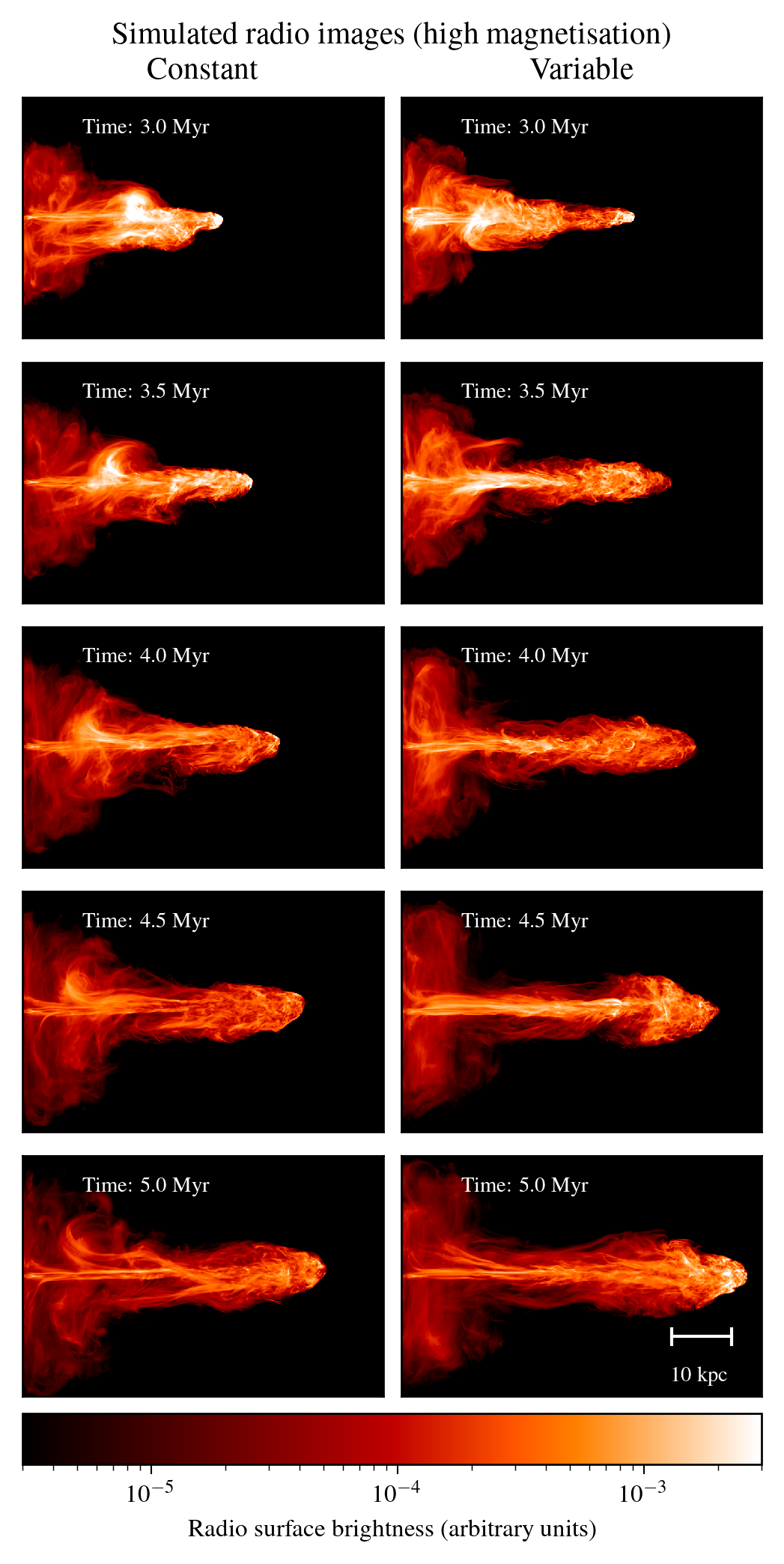}
    \caption{Comparison of simulated radio images of the constant and variable injection power, high magnetisation jets at various times in their evolution. At 3 Myr, the cocoon of the high magnetisation constant jet shows significant asymmetry around the jet head. At 3.5 and 4.0 Myr, this jet is misaligned with the central axis of the cocoon. The variable high magnetisation jet has a cocoon that is approximately symmetrical on large length scales throughout its evolution. The jet beam is not seen to deviate from its central position here, although the Doppler boosted example of this jet at 4 Myr in Figure~\ref{fig:seed_6_mag_010_compilation} shows significant misalignment in the approaching jet. Evidence of this misalignment is also seen in Figures~\ref{fig:broken_jets} and~\ref{fig:break_growth_seed6_010}. At 4.0 and 4.5 Myr, small brighter regions of the jet beam are visible at approximately halfway and three quarters of the way along the jet beam respectively. At 5.0 Myr the jet has a re-brightened hotspot.}
    \label{fig:time_comparison_010}
\end{figure}

The top left image of Figure~\ref{fig:time_comparison_010} shows a simulated radio image for this constant high magnetisation jet at 3 Myr. There is a bright hotspot region visible, around which the jet has recently carved out a narrow cylindrical volume of the surrounding medium. From this region, the strong one-sided backflow moves into the previously evacuated region below the jet beam, thus illuminating the strong asymmetrical shape of the cocoon near to the jet head. Should this process be occurring in sources, we would therefore expect to be able to see the asymmetrical cocoon shape, at least from some viewing angles, although projection effects may hide asymmetry should the asymmetry be aligned with our line of sight. If observed, large-scale asymmetries in cocoon shape near to the jet head which are misaligned with the current jet beam direction could be indicative of a recent change in direction of the jet beam. We show the simulated radio image at various times during the simulation (from the same line of sight) in Figure~\ref{fig:time_comparison_010}, showing that the asymmetrical cocoon shape feature does not persist throughout the jet's evolution, but that the off-centre position of the jet beam relative to the visible cocoon is a recurrent feature.

Figure~\ref{fig:time_comparison_010} also shows the simulated radio images of the high magnetisation variable jet. The departure of the jet beam from the central axis of the simulation is also seen in this case (see also Figure~\ref{fig:broken_jets}, discussed further in Section~\ref{sec:discontinuities_results}), however the strong example of the asymmetric excavation of the cocoon around the jet head (see Figures~\ref{fig:time_comparison_010} and~\ref{fig:Bx1_constant_010}) is not replicated in the variable case. Here the jet is unable to concentrate its ram pressure off axis for a period of $\sim1$ Myr, because the process is interrupted by discontinuities in the jet beam such as those discussed in Section~\ref{sec:discontinuities_results}.

Further examples of both kinds of asymmetrical behaviour can be found in the library of images we release with this paper. By extension, this library of images also exemplifies the consistent axisymmetric behaviour of the low magnetisation simulations.

\subsection{Discontinuities in the variable injection power high magnetisation jet beam}
\label{sec:discontinuities_results}
A `broken' morphology resulting from discontinuities in the jet beam is prevalent in the variable, high magnetisation jet. Firstly, we characterise the `broken' morphology. Secondly, we demonstrate the formation of the discontinuities in the high magnetisation variable jet by interactions between the magnetic fields and travelling shocks. Finally, we show that mixing prevents the formation of this broken morphology in the low magnetisation case.
\subsubsection{Characterising the `broken' morphology}
\label{sec:results_characterising_broken}
\begin{figure*}
    \centering
    \includegraphics[width=\linewidth]{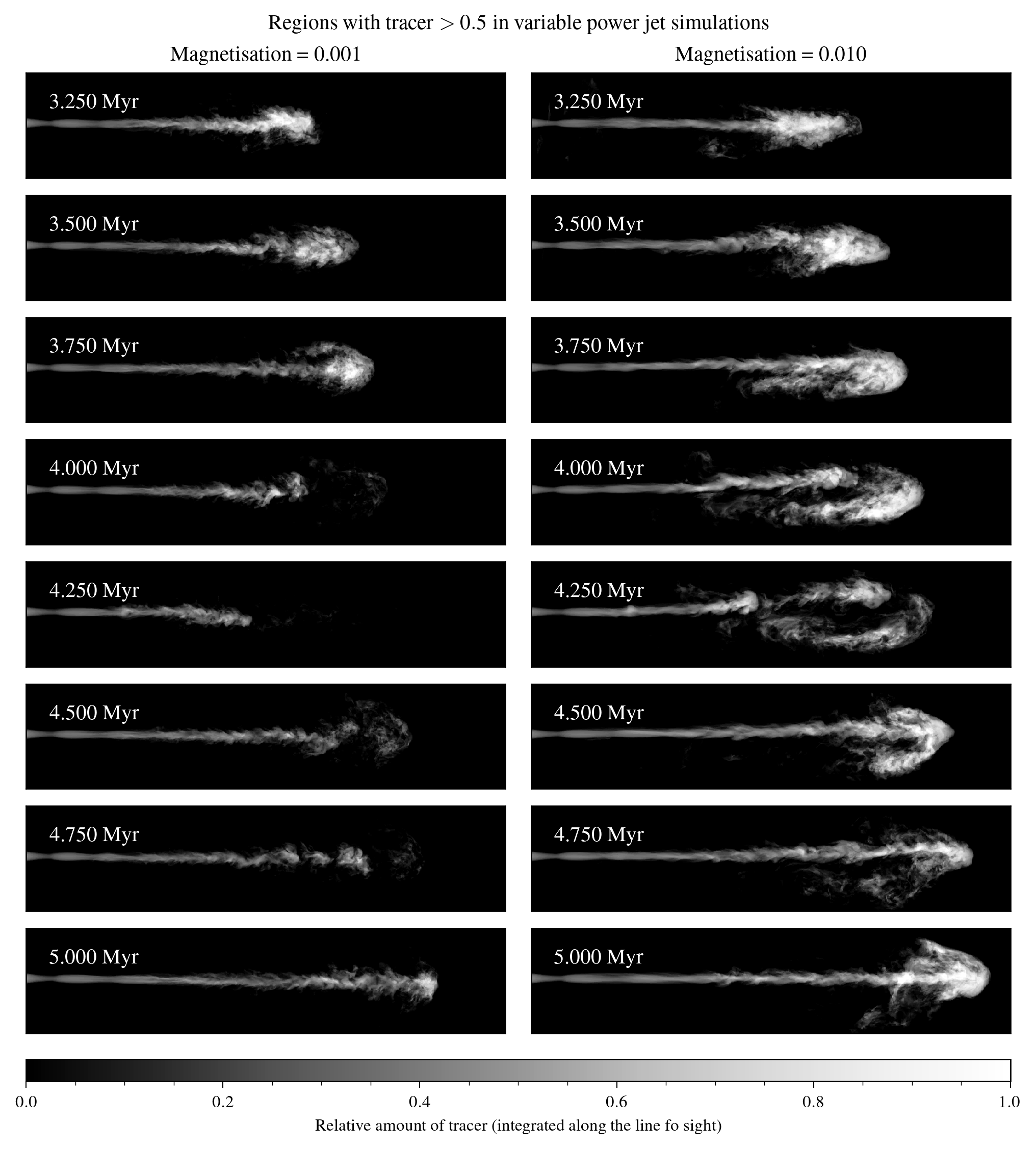}
    \caption{Tracer (fraction of material in a cell which entered the simulation through the jet nozzle) for the cells where the value of the tracer is above 0.5. The figure shows that the high magnetisation jet beam is more protected from mixing with the cocoon material and can sometimes form a discontinuous pattern of clumps of jet material travelling towards the jet head. The three-dimensional grid of tracer values is summed along the line of sight to produce these images and the section shown is 60 kpc long in the jet direction. White regions indicate that many cells along the line of sight have a value of tracer greater than 0.5, whilst black indicates that there are no cells along the line of sight where greater than half of the material entered the simulation through the jet nozzle.}
    \label{fig:broken_jets}
\end{figure*}
The variable, high magnetisation jet often displays discontinuities in the jet beam, similar to that shown in the tracer image in Figure~\ref{fig:summary_seed6_010} at 4 Myr and discussed in Section~\ref{sec:results_high_variable}. Whilst Figure~\ref{fig:summary_seed6_010} showed a central 2D slice through the simulation, Figure~\ref{fig:broken_jets} shows renderings of the jet tracer for both the low and high magnetisation variable power jet simulations between 3.25 and 5.00 Myr. Here we use the renderings of the jet tracer to gain a better understanding of the distribution in three dimensions. The jet tracer is the fraction of material in a cell which entered the simulation through the jet base. To create the rendered images, we first make a cut such that we only keep values at $\mathcal{C}>0.5$. This excludes diffuse emission from the lobes, allowing us to see the jet beam more clearly. We then sum the jet tracer value along the line of sight for any cells with $\mathcal{C}>0.5$.

Concentrating for now on the behaviour of the high magnetisation variable jet in Figure~\ref{fig:broken_jets}, it is clear that there are periods where the jet material forms clumps which travel towards the end of the jet with clear breaks between them. The clearest examples of this behaviour are at 4.00 and 4.25 Myr. We note that all images are made from the same viewing angle; all are 2D projections of a 3D shape, and so some may appear to have no break despite having one. The clumps generally appear to have a sharp cut off in concentration at their head, with a smoother tailing off in concentration at the end closer to the jet base. The sharp cut off at the head of the clumps suggests a mechanism that somehow cuts the jet, forming a definite boundary between the two regions. The clumps show evidence of disruption by hydrodynamic instabilities starting to mix the jet and non-jet material, but crucially they are able to survive for long enough to still be clumps, even once they enter the backflowing region (which is often one-sided as discussed in Section~\ref{sec:results_asymmetries}).

The discontinuous behaviour is in contrast to the constant power high magnetisation jet shown in Figure~\ref{fig:time_comparison_010}, which has a continuous jet beam at all times, extending from the base to the jet head, with no clear discontinuities. It is also in contrast to the behaviour of the low magnetisation variable power jet as we discuss in Section~\ref{sec:results_low_mag_variable_mixing}. This heavily suggests that the discontinuous `broken' morphology is a product of the interplay of variability and dynamically important magnetic fields (see Section~\ref{sec:results_how_breaks_formed}).
\subsubsection{Formation of discontinuities in the high magnetisation variable jet}
\label{sec:results_how_breaks_formed}
\begin{figure*}
    \centering
    \includegraphics[width=\linewidth]{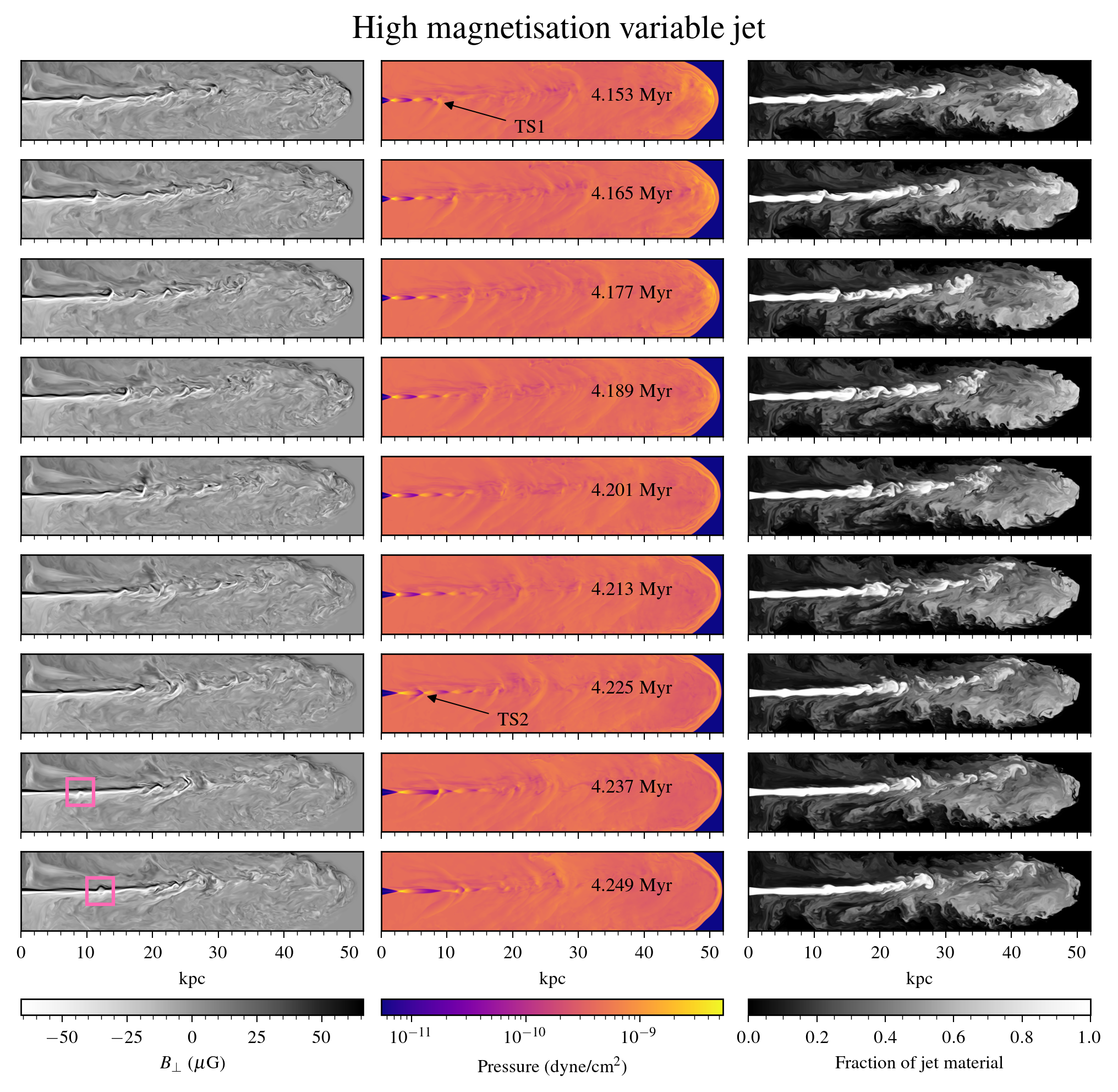}
    \caption{Broken jet structures form as a result of magnetic instabilities, which form due to propagating disturbances in the jet pressure. Slices through the variable high magnetisation jet simulation, showing the component of the magnetic field perpendicular to the slice ($B_\perp$, toroidal component at the jet base), pressure and the fraction of jet material (tracer). Limits on the pressure colour scale are chosen to accentuate the travelling shock structures, and do not accurately reflect the ambient medium pressure. TS1 and TS2 denote the approximate locations of travelling shocks 1 and 2 as described in the text.}
    \label{fig:break_growth_seed6_010}
\end{figure*}
Neither high magnetisation or variability alone is sufficient for the jet to form discontinuous structures -- rather it is the interaction between the two jet properties that leads to the broken morphologies. In non-magnetised jets, such as those we simulated previously~\citep{2026ElleySimulatingBrightening}, increases in jet power due to variability can cause travelling shocks to move from the jet base to the jet head, creating patches of bright emission and causing dramatic increases in the hotspot pressure and luminosity (see their figure 7). These increases in pressure at the jet head in non-magnetised jets are replicated in this work in the low magnetisation variable jet simulation (see Figure~\ref{fig:changing_hotspot_pressure}). In the high magnetisation case, the disruptions still travel from the jet base to the jet head, and can still cause a significant rise in pressure at the hotspot (as shown in Figure~\ref{fig:summary_seed6_010} and discussed in Section~\ref{sec:results_high_variable}). However, now the propagating disturbances can also cause sharp discontinuities in the jet material and are often accompanied by the growth of helical structures. In this section, we describe the formation of these discontinuities. Throughout, we refer to the structures as `kinked', although it should be noted that the shape alone does not definitively mean they must have been caused by the current-driven kink instability: we examine this possibility in Section~\ref{sec:discussion_formation}.

We begin by examining what happens in examples where the rate of change of the jet power increases. Two instances of this occur at 4.1 and 4.2 Myr (see Figure~\ref{fig:Lorentz_factors_timeseries}). Figure~\ref{fig:break_growth_seed6_010} shows $B_{\perp}$, pressure and jet tracer for a short section of time in the evolution of the variable high magnetisation jet, presented as a portion of a central slice through the simulation. The non-continuous changes in jet power cause travelling shocks to move along the jet -- the first of these (TS1) can be identified in the 4.153 Myr pressure map shown in Figure~\ref{fig:break_growth_seed6_010} at around 8 kpc from the base of the jet, and the second (TS2) at around 6 kpc from the jet base at 4.225 Myr. Both travelling shocks are labelled and can be seen to move steadily along the jet in the subsequent rows.

In the case of TS1, shown in the 4.153 Myr row, a kink has already developed in the magnetic field structure, coincident with the travelling shock. The tracer shows that this kink structure is affecting the spatial distribution of the jet material. Over the next two time snapshots shown, the kink structure grows and disrupts the jet, leading to a sharp break in the tracer distribution (shown in the third column of the figure). 

For TS2, Figure~\ref{fig:break_growth_seed6_010} shows the initial development of the kinked structure: at 4.225 Myr, the magnetic field morphology reflects the geometry of the travelling shock, whereas at 4.237 Myr, the magnetic field structure appears kinked. At 4.249 Myr the kinked structure has become more pronounced. This development history suggests that the instability is activated as a result of the interaction between the strong magnetic fields, the travelling shock and the recollimation shock. The location of the kinked structure can also be identified in the three dimensional render of the tracer at 4.25 Myr shown in Figure~\ref{fig:broken_jets} and in the emissivity map shown in the bottom left image of Figure~\ref{fig:seed_6_mag_010_compilation}, which is discussed further in Section~\ref{sec:predicted_images}.

\begin{figure*}
    \centering
    \includegraphics[width=\linewidth]{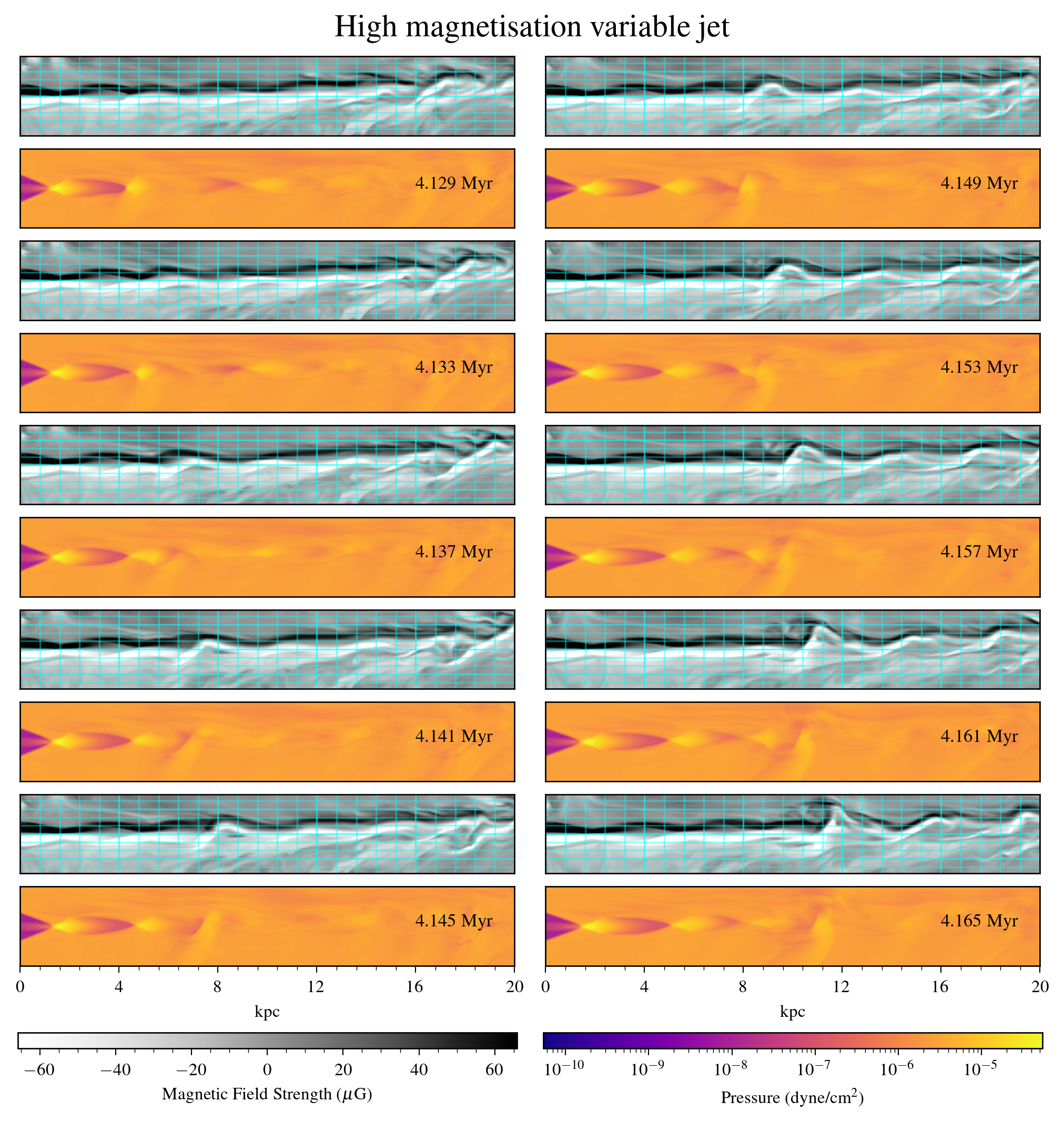}
    \caption{The growth of the kinked structures begins with small amplitude displacements from the approximately cylindrically symmetric geometry. These images show slices through the variable high magnetisation jet simulation, showing the component of the magnetic field perpendicular to the slice ($B_\perp$, toroidal component at the jet base) and the pressure. Limits on the pressure colour scale are chosen to accentuate the travelling shock structures. A grid is overlaid to aid in comparison between frames.}
    \label{fig:higher_time_res}
\end{figure*}

In Figure~\ref{fig:higher_time_res}, we show the magnetic field structure and pressure at higher time resolution and at a slightly earlier time in the jet's evolution than in Figure~\ref{fig:break_growth_seed6_010}. Figure~\ref{fig:higher_time_res} shows that the structures grow from small changes in the initial approximately cylindrically symmetric configuration, evolving into clear kinked structures over timescales of approximately 10-20 kyr.

Once the kinked structures have formed, grown and broken, they continue to move down the jet at the bulk flow velocity. In the case of TS1, Figure~\ref{fig:break_growth_seed6_010} shows that the in the region behind the travelling shock the process repeats with a second kinked structure appearing and breaking, leading to a complex tracer distribution in the section of jet behind the main break. 

This behaviour (the formation, growth and breaking of kinked structures) is in fact not limited to jet power increases and also occurs following a sharp {\em decrease} in the jet power. In Figure~\ref{fig:break_growth_following_decrease} we again show slices of $B_{\perp}$, pressure and jet tracer, but now showing the development of kinked structures for the time period following the sharp decrease in power at 4 Myr. Again, a disruption in the pressure travels along the jet, and with it the formation, growth and breaking of a kinked structure in the magnetic field structure. This break then becomes the discontinuous clumped structure seen at around 30 kpc in the first row (4.153 Myr) of Figure~\ref{fig:break_growth_seed6_010}.

\subsubsection{Why is this behaviour only seen in the high magnetisation case?}
\label{sec:results_low_mag_variable_mixing}
Previous works~\citep{Mukherjee2020SimulatingDynamics,2024RossiDifferentFlows} have highlighted that for jets simulated with a toroidal magnetic field, there is a  greater prevalence of KH instabilities in low magnetisation jets, with a more coherent magnetic field structure at high magnetisation. This preferential growth of K-H modes at low magnetisation is borne out in our simulations; we find that for our low magnetisation variable power jet, mixing via hydrodynamic instabilities prevents the survival of concentrated patches of jet material to large distances along the jet, such that the broken morphology cannot happen. In contrast, in the high magnetisation case, suppression of KH instabilities allows the preservation of discontinuous boundaries in jet material concentration and thus supports the formation of the clumped structures needed to produce the broken morphologies.

Returning to Figure~\ref{fig:broken_jets} and focusing on the low magnetisation jet, there are many times when the concentration of tracer drops below 0.5 significantly before the end of the jet (where $\mathcal{C}_j>0.5$ is the threshold we use to produce the rendered images, see Section~\ref{sec:results_characterising_broken}). Due to the high density contrast, it takes only a small amount of mixing with non-jet material by hydrodynamic instabilities to quickly reduce the fraction of jet material to below 0.5.
\begin{figure}
    \centering
    \includegraphics[width=\linewidth]{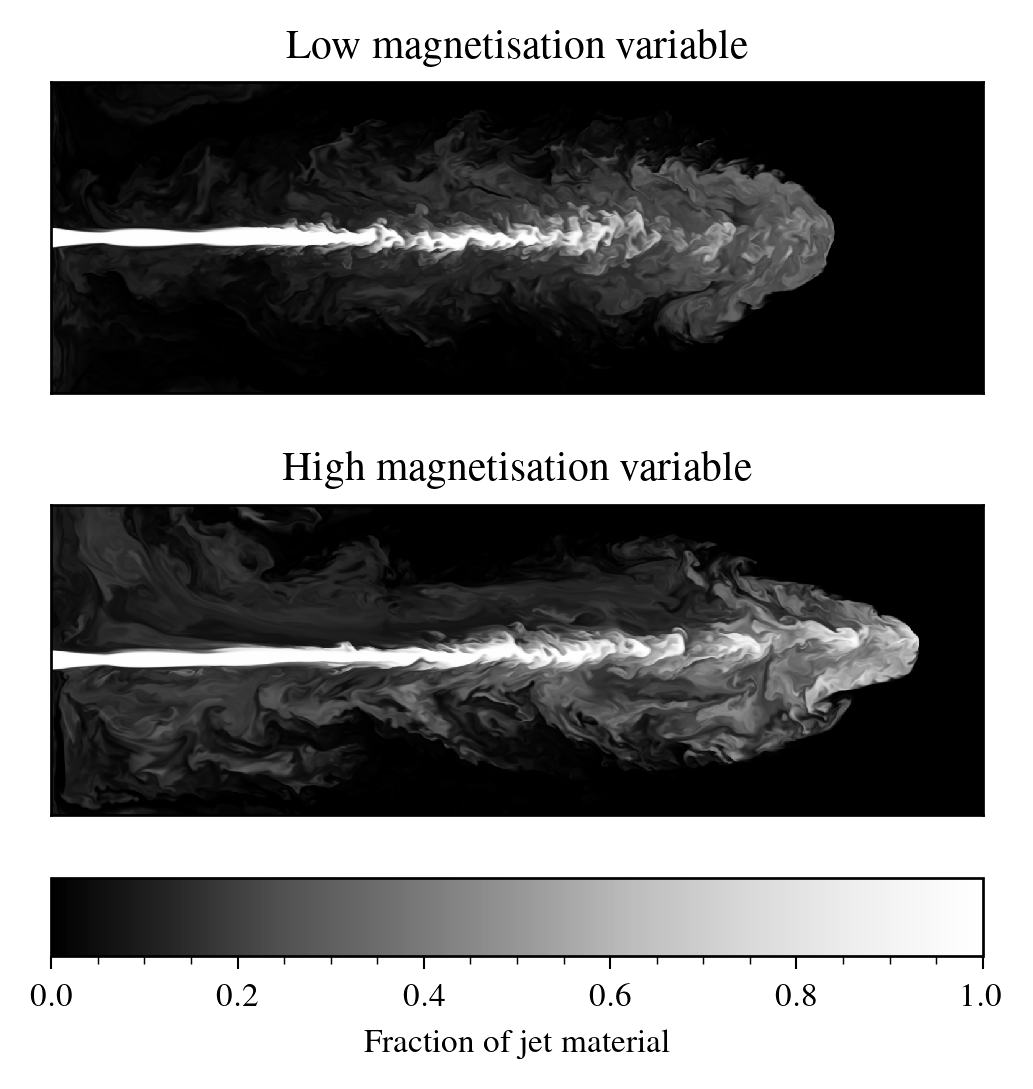}
    \caption{Central slices through the high and low magnetisation variable jets at 4.8 Myr showing the fraction of jet material (jet tracer). In the low magnetisation, variable power simulation, the jet beam is more heavily affected by mixing due to KH instabilities, whereas the high magnetisation, variable power simulation has higher fractions of jet material surviving to the jet head region. The slices shown are 60 kpc long in the jet direction.}
    \label{fig:mixing_illustration}
\end{figure}
 The role of hydrodynamic instabilities in mixing the jet material with non-jet material is illustrated most clearly by central 2D slices of the jet tracer, which we show for the two variable jets in Figure~\ref{fig:mixing_illustration}. In the low magnetisation case (top panel), the jet tracer has dropped to low values by the time material reaches the jet head, with a smooth distribution on large length scales outside of the jet beam. Turbulent structures (indicative of mixing by KH instabilities) are visible in the tracer distribution with characteristic sizes of approximately 1-2 kpc. Furthermore, these turbulent structures dominate the small scale structure of the jet head region in the simulated radio images of the low magnetisation jets shown in Figure~\ref{fig:time_comparison_001}.

At high magnetisation, the jet beam is altogether more coherent. At all times shown in Figure~\ref{fig:broken_jets}, the high magnetisation jet has significant volumes close to the jet head that are populated with material with a tracer value of over 0.5, indicating that mixing via KH instabilities is not efficient enough to fully erode the clumped structures created by the variability and magnetic fields. Further to this, the simulated radio images of the high magnetisation jet simulations often show long-range ($\gtrsim 5$ kpc) filamentary structures near to the jet head (the high resolution images in Figure~\ref{fig:seed_6_mag_010_compilation} show this most clearly), similar to those seen in previous works simulating magnetised jets~\citep[e.g][]{2024UpretiBridgingPolarisation}. Whilst the high magnetisation simulations do show evidence of turbulence in this jet head region, its impact is visibly reduced compared to the low magnetisation simulations (compare for example Figures~\ref{fig:seed_6_mag_010_compilation} and~\ref{fig:seed_6_mag_001_compilation}).

\section{Discussion of beam stability}
\label{sec:discussion_stability}
We described the formation of discontinuities in the high magnetisation variable injection power jet simulation in Section~\ref{sec:discontinuities_results}. Here we discuss a possible formation mechanism and use an analytic description of the injected equilibrium to motivate the expectation that development of wavelike and kinked jet behaviour should occur at the smallest possible wavelengths. We begin by confirming that the injected jet profile is in a stable equilibrium across the jet cross section, before estimating the wavelength of waves that could form assuming the jet remains in an analogous equilibrium state. In fact, following injection, the jet undergoes a series of recollimation shocks, due to the greater pressure in the cocoon than within the beam itself. These recollimation shocks complicate the radial profiles, and form the same shape as an $m=0$ pinch instability would, however, here the pinched shape is caused by the cocoon pressure, rather than an instability due to the structure of the injected beam itself. To proceed with the analysis, we assume that the beam returns to a state with an at least qualitatively analogous profile to that which is injected. Focusing on the images in Figure~\ref{fig:higher_time_res}, it can be seen that in the region between successive recollimation shocks, the jet returns to a structure with a magnetic field that peaks at a given radius and with a pressure profile that is highest in the centre of the beam, as it is at injection. We focus on describing the asymmetric behaviour seen in our simulations, and make assumptions in the analysis that are suitable for our resolution; for example, we assume any perturbation has a constant magnitude across all radii and we do not consider cross terms between forces and displacements in $\hat{r}$, $\hat{\phi}$ and $\hat{z}$ directions. The cocoon-pressure driven $m=0$ deformations of the beam are present also in the constant power case -- here we focus on identifying a possible mechanism leading to asymmetric behaviour in the variable case, and as such look predominantly at the $m=1$ modes. As a result of the above, the picture we build in this section of a jet beam being taken out of a stable equilibrium and the subsequent formation of wavelike and kink instabilities is idealised. The processes that occurring in the interaction between the travelling shocks and the magnetic field structures and lead to the formation of kinked structures as discussed in Section~\ref{sec:results_how_breaks_formed} are complex. Whilst we neglect some complexities, our approach here allows us to provide a possible interpretation of the behaviour seen in the simulations. Previous works have suggested the current-driven kink instability as the driver of similar structures in X-ray binary~\citep{2022LopexMiralles3DBinaries} and AGN~\citep{2024RossiDifferentFlows} jet simulations.

\subsection{Stability of the injected equilibrium}
\label{sec:discussion_stability_inner}
We inject a magnetic field and pressure structure which is chosen to be in equilibrium (see Sections~\ref{sec:methods_equilibria} and~\ref{sec:methods_pressure_profile}, in particular Equations~\ref{eqn:equilibrium_condition} and~\ref{eqn:specific_equilibrium}). We demonstrate that our chosen injected equilibrium is stable using two approaches -- firstly, we refer to the behaviour of the constant power simulations, specifically Figure~\ref{fig:Bx1_constant_010}. Taking the times from 2.0 Myr onwards, when the jet length has grown to a significant fraction of the size of the simulation domain, whilst the jet bends on long length scales, the section of the jet immediately following injection is approximately straight, with its behaviour dominated by axially symmetrical recollimation shocks. 

Secondly, we demonstrate analytically that the force on a fluid element displaced slightly from its equilibrium position is a restoring force. The force (per unit volume) on a fluid element due to a displacement $\vec{\xi}$ from equilibrium is~\citep[e.g.][]{Goedbloed_Keppens_Poedts_2019}
\begin{multline}
        \vec{F}(\vec{\xi}) =-\vec{\nabla}(-\gamma_{\rm{ad}} p\vec{\nabla}\cdot\vec{\xi}-\vec{\xi}\cdot\vec{\nabla}p)\\
    -\vec{B}\times(\vec{\nabla}\times(\vec{\nabla}\times(\vec{\xi}\times\vec{B})))\\
    +(\vec{\nabla}\times \vec{B})\times(\vec{\nabla}\times(\vec{\xi}\times\vec{B})) \\
    \equiv \vec{F}_p(\vec{\xi}) + \vec{F}_{B,1}(\vec{\xi}) + \vec{F}_{B,2}(\vec{\xi})\equiv -(F_p + F_{B,1} -F_{B,2})\vec{\xi}
\label{eqn:force}
\end{multline}
where $\gamma_{\rm{ad}}$ is the adiabatic index and we include the formulation with scalar factors $F_p$ , $F_{B,1}$, and $F_{B,2}$ for ease of comparison between the competing terms in Section~\ref{sec:discussion_formation}. All forces are given in the frame comoving with the jet velocity -- we have omitted the primes on the forces for clarity. We consider a small radial displacement
\begin{equation}
    \vec{\xi}(r,\phi,z',t) = \xi e^{i(m\phi+k'z'-\omega t)}\hat{r},
\end{equation}
and only keep forces in the $\hat{r}$ direction throughout this analysis. We assume that all material along a radial path is displaced in phase, the equivalent to setting $l=0$ in a displacement of the form $e^{i(m\phi+k'z'+lr-\omega t)}$. This can equally be thought of as investigating only surface modes~\citep[see][for a description of possible structures in jets]{2000HardeeOnJets}. This is a reasonable approximation to make for our aim of describing the behaviour in our simulations, as the radius of the inner region of the jet, $a$, is spanned by only 5 cells, such that wavelengths $\lambda\lesssim a$ will not be well resolved. In the region $r<a$, the first term of equation~\ref{eqn:force}:
\begin{equation}
    \vec{F}_p(\vec{\xi})
    =-\left(\gamma_{\rm{ad}}\left(\frac{p_a}{r^2}+\frac{B_\phi'^2}{4\pi r^2}+\frac{B_\phi'^2}{4\pi a^2}\right)+\frac{2B_\phi'^2}{4\pi a^2}\right)\vec{\xi}
    \label{eqn:first_force_term}
\end{equation}
such that the thermal pressure field we choose provides a restoring force to small perturbations. The second term of Equation~\ref{eqn:force}
\begin{equation}
    \vec{F}_{B,1}(\vec{\xi})=\frac{B'^2_\phi}{4\pi a^2}(1-m^2)\vec{\xi}
    \label{eqn:second_force_term}
\end{equation}
which is zero for the $m=1$ kink mode. The third term of Equation~\ref{eqn:force} is given by
\begin{equation}
    \vec{F}_{B,2}(\vec{\xi})
    =\frac{2B_\phi'^2}{4\pi a^2}\vec{\xi}
    \label{eqn:third_force_term}
\end{equation}
The magnetic forces can be readily verified against those found via application of Equation 3.22 of~\cite{1998BegelmanInstabilityPlerions} in the case of $m=0$. There is a cancellation between the result of Equation~\ref{eqn:third_force_term} and the final term of Equation~\ref{eqn:first_force_term}, such that the full force is
\begin{equation}
    \vec{F}(\vec{\xi})=-\gamma_{\rm{ad}}\left(\frac{p_a}{r^2}+\frac{B_\phi'^2}{4\pi r^2}+\frac{B_\phi'^2}{4\pi a^2}\right)\vec{\xi}.
    \label{eqn:restoring_force}
\end{equation}
This is always a restoring force for the field we inject, confirming that the injected field is a stable equilibrium. We present an analysis of the stability of the field in the region $r\geq a$ in Appendix~\ref{sec:appendix_stability_outer}, showing that the forces on this region are small compared to the region $r<a$.

Considering briefly the $m=0$ pinch mode, we note that only Equation~\ref{eqn:second_force_term} has dependence on $m$ and thus the $\vec{F}_{B,1}(\vec{\xi})$ provides a positive force in the $m=0$ case. This force is however still not enough to overcome the stabilising forces from the pressure term given by Equation~\ref{eqn:first_force_term}. Given that we seek to describe the asymmetric behaviour found in the simulations, from here on we consider only the $m=1$ mode.

In the above analysis, we have focused on a description of the behaviour in our simulated jets, and as such have assumed $l=0$ and focused only on radial forces resulting from a radial displacement. For a description of the behaviour on scales smaller than our grid resolution, we refer the reader to \cite{1998BegelmanInstabilityPlerions}, who find that small wavelength modes can be unstable with the fastest growing mode of the $m=1$ instability
occurring for $l/k\ll1$, such that the wavelength along the $\hat{z}$ direction is much smaller than the wavelength in the $\hat{r}$ direction. These small wavelength modes grow on short timescales $<0.01 Myr$ and \cite{1998BegelmanInstabilityPlerions} argue that a finite resistivity is likely to dissipate magnetic energy via reconnection of field lines. This subgrid effect is not captured in our simulations and could lead to additional particle acceleration along the jet~\citep[e.g.][]{2020DavelaarParticleJets}. On the other hand, \cite{1994IstominStabilityJet} suggest that modes with wavelengths on the order of the jet radius are those most likely to grow to the largest amplitudes, further supporting our choice to focus on $l=0$ modes. Simulations of more idealised jet sections at higher resolutions~\citep[e.g.][]{2009MizunoThreeColumn, 2024MussoEvolutionJets} are in general more suited to the analysis of the relative growth of smaller wavelength modes and their effects, and in general recover clear growth of modes with wavelengths comparable to the jet radius. Overall these results suggest that whilst our simulations do not resolve the growth of short-wavelength modes, the most likely effect of resolving them would be additional particle acceleration, with limited effects on the large-scale morphology.

\subsection{Wavelike modes in stable equilibria}
The magnitude of the restoring force is directly proportional to the displacement and as a consequence we expect it to act in a way analogous to a simple harmonic oscillator, admitting wavelike behaviour at a single frequency.  For illustrative purposes, we show that for reasonable values taken from the simulation domains, wavelike modes with wavelengths of $\approx 1$ kpc can form. We assume the same form of equilibrium as above, but use values from the simulations at around 10 kpc along the jet for the pressure, magnetic field strengths and densities.

We use the wave equation
\begin{equation}
    \frac{\partial^2\vec{\xi}}{\partial t^2} = v_A^2\frac{\partial^2\vec{\xi}}{\partial z'^2}
\end{equation}
and Equation~\ref{eqn:restoring_force} evaluated at $r=a$ to get an estimate for the wavenumber $k'$
\begin{equation}
    k'=\frac{1}{av_A}\sqrt{\frac{\gamma_{\rm{ad}}}{\rho'}\left(p_a+\frac{B_\phi'^2}{2\pi}\right)}=\frac{1}{0.2v}\sqrt{\frac{4}{3}\frac{1}{\rho'}\left(p_a+\frac{B_\phi'^2}{2\pi}\right)}\rm{\,kpc}^{-1},
\end{equation}
where $v_A$ is the Alfvèn speed. In Table~\ref{tab:wavelength_estimates}, we take representative values from the slices through the high magnetisation simulations (focusing on the jet edge) and use them to estimate the wavelengths that can form. We estimate the Alfvèn speed using
\begin{equation}
    v_A=\frac{c}{\sqrt{1+\frac{4\pi(e+p)}{B^2}}}\approx\frac{c}{\sqrt{2+\frac{4\pi p}{B^2}}},
\end{equation}
taking the energy density in relativistic particles to be $e\approx u_B= \frac{B'^2}{4\pi}$, corresponding to equipartition with the energy density in the magnetic fields (more conservative here than considering a fraction of equipartition). For the range of values in Table~\ref{tab:wavelength_estimates}, this gives an Alfvèn speed of $v_a \approx 0.4-0.7c$. For simplicity, to calculate the wavenumbers and wavelengths in Table~\ref{tab:wavelength_estimates} we assume $v_A\approx c$ and use a bulk Lorentz factor of $\Gamma_j=4$ for all transformations. Using a smaller value for $v_a$ would slightly increase the wavenumbers and decrease the wavelengths, but does not significantly affect our results. In all cases, the wavelike behaviour resulting from the perturbation of the equilibrium state has a predicted wavelength in the lab frame that is too small to be resolved by our simulations. Stable, short wavelength modes such as these are unlikely to have significant effects on the overall behaviour of the fluid on the large scales probed by our simulations.
\begin{table*}
    \centering
    \begin{tabular}{c|c|c|c|c|c|c|c}
    \hline
         Description & $p_a \rm{\,(dyne\,cm}^{-2})$ & $B_\phi'\,(\rm{\mu G})$ & $\rho_j\,(\rm{g\,cm}^{-3})$ & $\rho_j'\,(\rm{g\,cm}^{-3})$ & $k'\,(\rm{kpc}^{-1})$ & $\lambda'\,(\rm{kpc})$ & $\lambda\,(\rm{kpc})$\\\hline
         Edge of jet beam & $10^{-10}$& $50$& $10^{-30}$ & $2.5\times10^{-30}$ & 2.7 & 0.4 & 0.1\\
         High pressure region, recollimation shock & $10^{-9}$& $50$& $10^{-29}$ & $2.5\times10^{-29}$ & 1.4 & 0.7& 0.2\\
         Low pressure region, recollimation shock & $10^{-11}$& $50$& $10^{-31}$ & $2.5\times10^{-31}$ & 7.8 & 0.1&0.03\\
         \hline
    \end{tabular}
    \caption{Table showing estimated wavelengths of waves travelling at $~\approx c$ in three representative sections of the jet beam. The first row is chosen to be representative of a section of the jet beam which is not subject to strong effects from recollimation shocks. The second and third rows instead focus on extreme regions within the jet. All rows suggest a wavelength which is small compared to the width of the jet, suggesting that these are unlikely to form.}
    \label{tab:wavelength_estimates}
\end{table*}

\subsection{Formation of helical structures and localised kink structures}
\label{sec:discussion_formation}
During the above analysis we assumed that the jet remains in a stable equilibrium. At many times in the high magnetisation simulations, this is a reasonable assumption -- particularly near to the jet base, the jet beam often remains straight on scales of $\approx 10$ kpc, without signs of asymmetric oscillations. At other times, we see that the jet displays asymmetric wavelike structures with $\lambda\approx3$ kpc and the growth of kinked structures with a typical size of $\approx1-2$ kpc (see the magnetic field slices in Figure~\ref{fig:break_growth_seed6_010}). The two behaviours happen in both the constant and variable power jets, but they are more frequent in the variable case and happen close to the jet base, in a way not seen in the constant power jets. We suggest that the formation of these wavelike and kinked structures may happen as a result of a departure from the stable equilibrium due to interactions between the jet beam and the cocoon. In particular, we show that in the idealised case, compression along the jet beam may destabilise small sections of the jet to the $m=1$ kink instability. In fact, as previously discussed, the jet may also be expected to become unstable to the $m=0$ instability. We do not discuss this here, as the symmetric behaviour in the simulations is dominated by recollimation shocks.

Of key importance to the injected equilibrium and its stability is the balance between the pressure gradient and the radial forces created by the magnetic field -- the potentially destabilising force provided by $F_{B,2}$ is balanced by part of $F_P$. Travelling shocks caused by changes in the jet power (here implemented through changes in jet speed) can compress the plasma and toroidal magnetic fields along the jet beam direction, whilst recollimation shocks compress the jet beam radially. In the case of axisymmetric radial compression of the jet, the toroidal magnetic field strength $B'_\phi$ increases such that $B'_\phi\propto1/r_j$. To obtain a naive estimate of the pressure increase, we consider the thermal energy of the jet remaining constant, but being compressed into a smaller volume, suggesting a scaling of $p_j\propto1/r_j^2$. Thus the equilibrium condition~\ref{eqn:specific_equilibrium} is still fulfilled by the compressed structure, although it may be unlikely to be in equilibrium with its surroundings. On the other hand, if a section of jet is compressed along the jet direction, we still have $B'_\phi\propto1/\Delta z$ but the pressure instead scales as $p_j\propto1/\Delta z$, where $\Delta z$ is the length of the compressed section. The equilibrium condition~\ref{eqn:specific_equilibrium} is then broken and the jet will contract radially. Superimposed on this behaviour may be non-axisymmetric behaviour such as the growth of a kink instability. To gain a qualitative description, we assume that the jet remains close to the equilibrium described previously, considering the effects of small changes in the field structure.

A key feature of the derivation of Equation~\ref{eqn:restoring_force} is that Equation~\ref{eqn:first_force_term} depends only on the pressure, $F_P\propto p$: the appearance of $B_\phi'$ on the right hand side of Equation~\ref{eqn:first_force_term} follows directly from having specified the pressure gradient to match the magnetic field in Equation~\ref{eqn:pressure_profile} (necessary to obtain an initial equilibrium). On the other hand, $F_{B,2}\propto B_\phi'^2$ in its own right. As a result, in the case of compression along the jet beam direction where $B_\phi'^2$ grows faster than the pressure, $F_{B,2}$ can feasibly grow to  $F_{B,2}\lesssim F_P$, leading to a weaker restoring force or even such that $F_{B,2}>F_P$, leading to a kink- (and pinch-) unstable region of the jet (although the $m=0$ behaviour is dominated by the recollimation shocks).

A reduced restoring force would increase the wavelength of wavelike modes forming in the jet beam (together with a reduced frequency of oscillation). The smallest wavelengths that should be resolved in the simulation are around 0.4 kpc in wavelength, assuming 10 cells are needed per wavelength. The jet width is likely a more stringent limit -- for a structure which is 0.4 kpc across (measuring between the peak values of the magnetic field), to start seeing wavelike behaviour in the simulations, the wavelength should be $\gtrsim1$ kpc. If this threshold is reached, a small perturbation to the jet's shape (such as that seen in the first panel of Figure~\ref{fig:higher_time_res}) due to interactions between the jet beam and the cocoon, such as interactions with the backflow, will cause wavelike behaviour. A wavelike pattern in the magnetic field structure with a high enough amplitude could then feasibly enter a kink unstable regime, due to the changes in the relative magnetic field strengths on each side of the beam. This could break the beam up into clumped sections. In the first few rows of Figure~\ref{fig:break_growth_seed6_010}, the section of jet initially between 10 and 20 kpc shows wavelike behaviour with a characteristic wavelength of $\lambda\approx3-4$ kpc which grows in amplitude. This wavelength is close to the spacing between the recollimation shocks, which could imply a physical relation between the two phenomena. The structures appear to have broken at 4.177 Myr, although if we only look at these 2D slices, it is possible the structure is helical and instead moves into and out of the plane. The morphology of the tracer render at 4.250 Myr in Figure~\ref{fig:broken_jets} confirms that significant breaks have occurred within this section (in addition to the two clearest breaks).

Alternatively, if the force on a displaced fluid element increases with displacement $\xi$ -- for example, due to a significant compression along the jet beam direction -- the period of wavelike behaviour can be completely bypassed. In this case the jet would become kink unstable in the region of the compression. In the last two rows of Figure~\ref{fig:break_growth_seed6_010} we see the very localised growth of a kinked shape with a smaller characteristic size at a distance of around 10 kpc along the jet (marked in Figure~\ref{fig:break_growth_seed6_010} by the pink boxes). In contrast to the onset of wavelike behaviour, which affects $\approx 10$ kpc sections of the jet beam, the growth of localised kinks in the jet affects only a very small section of the jet beam $\approx 1-2$ kpc in length.

Whilst we have suggested a specific way in which a jet injected in a stable equilibrium can become subject to wavelike behaviours on kpc scales and to the $m=1$ kink instability, we note that the situation is much more complex, both in terms of interactions between the jet and cocoon, and the non-linear development of instabilities past their initial onset. The use of simulation-based techniques is therefore crucial to understand the expected outcomes. 
\section{Predicted Images and Observational Implications}
\label{sec:predicted_images}
\subsection{Appearance of broken jet structures in simulated radio images}
\label{sec:appearance_broken_structures}
\begin{figure*}
    \centering
    \includegraphics[width=\linewidth]{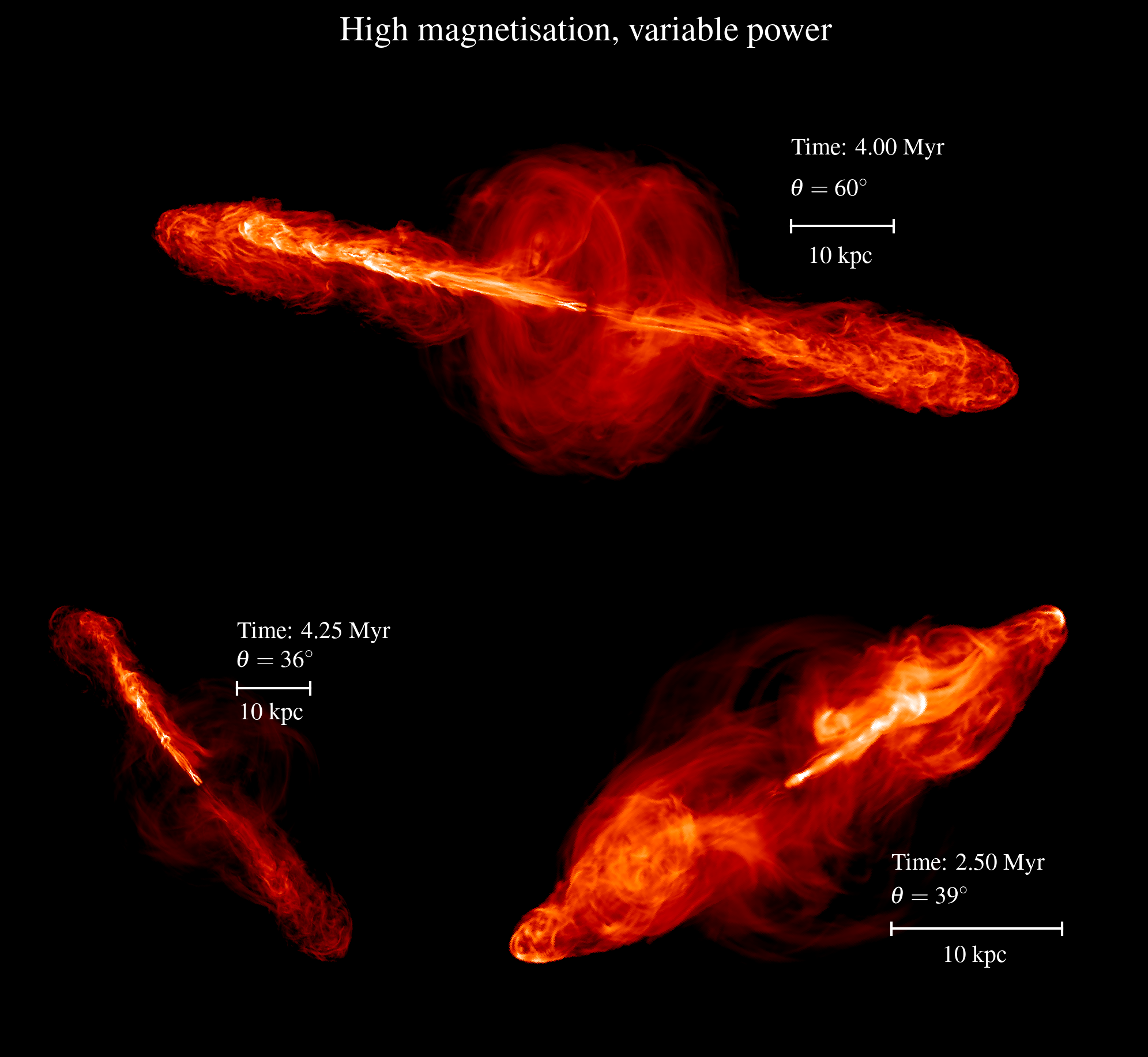}
    \caption{Three illustrative examples of simulated radio images for the high magnetisation, variable power jet at different times in its evolution. All images are made with a dynamic range of 3 dex in brightness, however the limits of the brightness scale vary between images to account for differences in overall brightness due to age and boosting. The spatial scale (which reflects the distance projected onto the plane of the sky, rather than the physical jet length) for each image is indicated by the corresponding scale bar; images are presented at different spatial scales to facilitate seeing various details and features in each. The corresponding times and views for the low magnetisation variable jet simulation are presented in Figure~\ref{fig:seed_6_mag_001_compilation}. $\theta$ gives the angle between the approaching (brighter) jet and the line of sight.}
    \label{fig:seed_6_mag_010_compilation}
\end{figure*}

We have shown that broken jet morphologies can form in a variable injection power jetted source with a dynamically significant magnetic field. We turn now to asking whether this phenomenon would be observable. In Figure~\ref{fig:seed_6_mag_010_compilation} we show simulated radio images of the high magnetisation variable jet simulation (calculated following Section~\ref{sec:rendering}) for three illustrative time stamps and angles. The top image ($4$ Myr) shows the approaching jet beam (left of image) as a brightly emitting region extending for the majority, but not quite all, of the length of cocoon. In Figure~\ref{fig:broken_jets}, we showed that the jet tracer showed a clear discontinuity in the corresponding snapshot and at the same location in the jet. The receding jet, however, is dimmer and lower contrast, due to Doppler de-boosting, and thus shows no clear signature of this behaviour, despite being a reflection of the same simulation data.

The bottom-left image of Figure~\ref{fig:seed_6_mag_010_compilation} shows the high magnetisation variable jet at slightly later time of 4.25 Myr. We opt for a different line of sight ($\theta=36^\circ$), more closely aligned with the jet axis, for this image. The small kinked structure visible at approximately 12 kpc along the approaching jet in Figure~\ref{fig:break_growth_seed6_010} is visible here as a small bright region approximately 1/3 of the way along the jet length. There is also a bright region of emission approximately halfway along the approaching jet beam, which maps directly to a region of high tracer with an abrupt discontinuity, as can be seen in the 4.25 Myr row of Figure~\ref{fig:broken_jets} and the 4.249 Myr row of Figure~\ref{fig:break_growth_seed6_010}. In general, the distinct parcels of material seen in the tracer images appear more diffuse in simulated radio images; however, the overall path of the more brightly emitting jet material follows that of the tracer, such that the large scale asymmetrical behaviour of the jet beam -- having moved away from the central axis -- remains fairly clear.

The bottom-right image of Figure~\ref{fig:seed_6_mag_010_compilation} shows a helical structure formed behind a break in the high magnetisation variable jet at 2.5 Myr. We expect such well-defined helical structures to be a necessarily short-lived phase in the evolution, marking the transition from a linear to a non-linear growth phase. In other words, to reach a helical geometry, the instability must have been growing exponentially and must be about to break the jet apart in a complex manner. Figure~\ref{fig:break_growth_seed6_010} illustrates the short-lived nature of this phase, showing the formation of such a structure at 4.165 Myr and at 4.213 Myr. In both cases, the structure is disrupted over the next one or two rows, corresponding to timescales of $\approx0.01$ Myr. Figure~\ref{fig:higher_time_res} further illustrates this growth of amplitude of a similar structure from a small amplitude to a non-linear phase with an amplitude much greater than the jet width, which then snaps. In contrast, the resulting broken morphologies, with the brightest sections of jets not reaching to the end of the cocoon are longer lasting in nature, depending instead on the bulk speed of the jet material and therefore able to persist for approximately 0.1 Myr. From an observational point of view, this suggests that, should a section of the jet become unstable to the kink instability, we should expect it to move through an observable helical phase, but that we shouldn't expect this to last for a large fraction of the jet's lifetime. We should therefore only expect to see it in a small subset of sources, even if the overall phenomenon were a ubiquitous process in radio galaxies. We note that we captured the three-dimensional data required to produce the emissivity maps at a frequency of 0.25 Myr for the variable jets; in these snapshots, the example of the helical structure given in Figure~\ref{fig:seed_6_mag_010_compilation} at 2.5 Myr is the only clear example we see in the simulated images. The broken morphologies, on the other hand should be present in many more observations.

\subsection{Viewing angle effects}
\label{sec:viewing_angle_effects}
Doppler boosting and projection effects can have significant consequences on the observed morphology of radio galaxies and on the appearance of the features we have shown in our simulated radio galaxies. We illustrate these effects using our constant power jet with high magnetisation. We split the analysis into two parts: firstly we discuss the effects of rotation of the jet about the jet axis, where projection effects can make asymmetries more or less pronounced. Secondly, we rotate the jet such that the angle between the line of sight and the jet axis varies. Here, the predominant effect is the Doppler boosting of the material travelling relativistically along the jet beam.

\begin{figure}
    \centering
    \includegraphics[width=\linewidth]{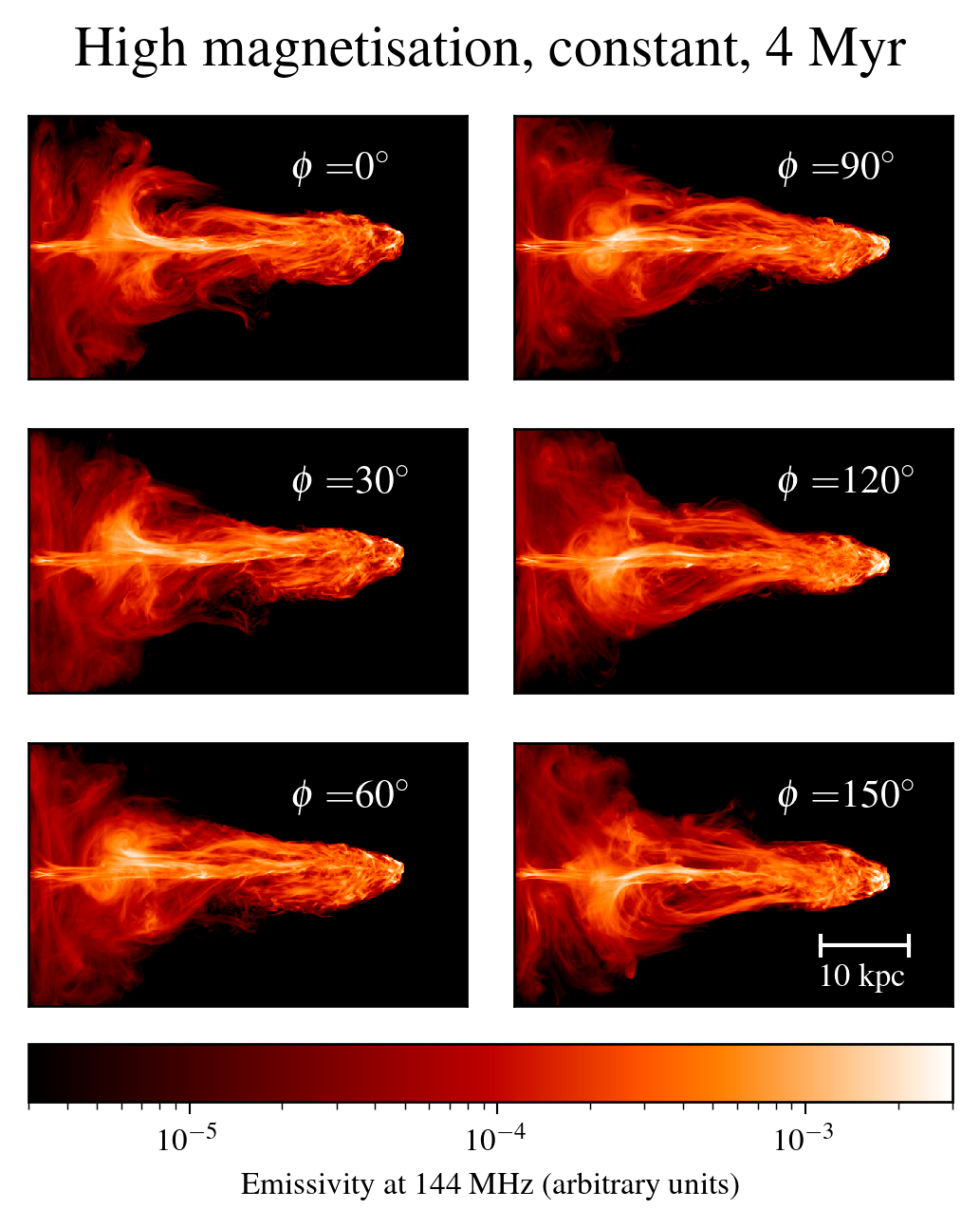}
    \caption{high magnetisation, constant power jet at 4 Myr as viewed from varying azimuthal angles (varying $\phi$), showing that the asymmetric cocoon structure near to the jet head is much more prominent from some viewing angles (e.g. $\phi=0^\circ$), whilst at others the cocoon shape appears symmetric(e.g. $\phi=150^\circ$). Additionally, the brightness of the hotspot varies with viewing angle, due to Doppler boosting of the asymmetric backflow near to the jet head.}
    \label{fig:rotation_comparison}
\end{figure}

Figure~\ref{fig:rotation_comparison} shows the constant power jet with $\sigma_B=0.01$ at 4 Myr from different perspectives, but all with the jet axis aligned with the plane of the sky. From some azimuthal angles, the cocoon shape at the jet head is clearly asymmetrical. From others, the jet appears misaligned with the central axis of the cocoon. In summary, whilst some aspect of large-scale asymmetry is present in most of the panels shown, there is a large amount of variety in both the extent and nature of the asymmetry that would be inferred based on observations from various viewing angles.

\begin{figure}
    \centering
    \includegraphics[width=\linewidth]{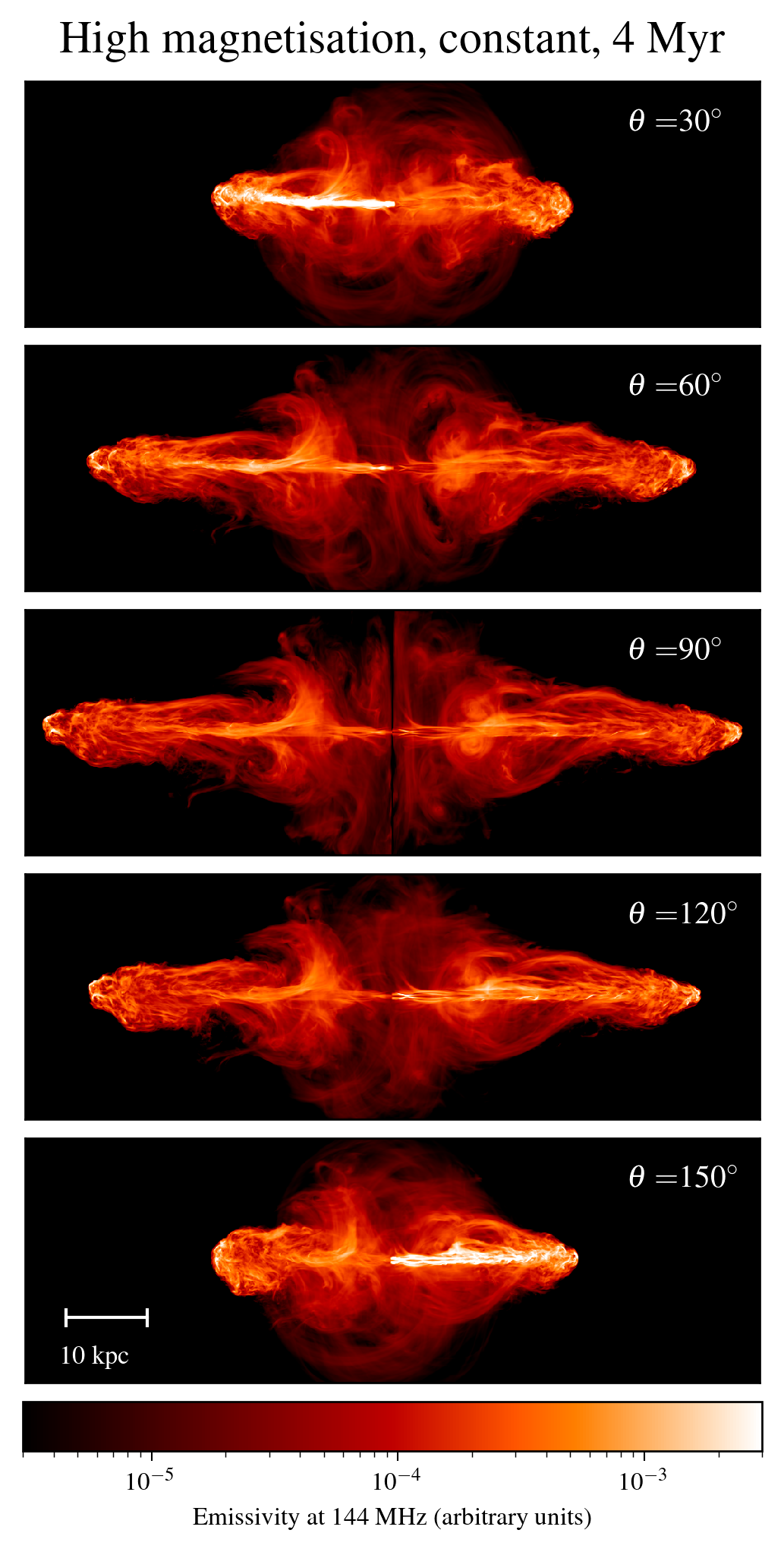}
    \caption{High magnetisation, constant power jet at 4 Myr. The angle between the left hand jet and the line of sight is given in the top right of each image. In the first row, the left hand jet is approaching the observer. As we move through the images, this left hand jet moves away from the observer, through the plane of the sky in the third image and then continues until the right hand jet is approaching (and at an angle of 30 degrees to the line of sight) in the fifth image.}
    \label{fig:boosting_comparison}
\end{figure}

In Figure~\ref{fig:boosting_comparison}, we move to showing the same simulation data at the same epoch as in Figure~\ref{fig:rotation_comparison}, but this time rotated such that the angle between the jet beam and the line of sight changes.  Figure~\ref{fig:boosting_comparison} shows significant effects from Doppler boosting of the jet beam and backflows. Differential Doppler boosting results in clear changes in the relative brightness of the jet beam and cocoon in the images, which in turn affects how clearly we can see that the jet has been bent away from the central axis of the simulation. The central panel of Figure~\ref{fig:boosting_comparison}, where the jet is aligned with the plane of the sky, illustrates the approach taken in the creation of the two-sided images. We take the simulation data, reflect it through the $z=0$ plane and rotate it about the z axis by 90 degrees. When the left hand jet beam approaches the observer and is Doppler-boosted, the slight bend in the left hand jet becomes much more prominent and it becomes much clearer visually that the jet beam is bent towards one side of the inflated lobe. When the right hand jet is approaching, the Doppler-boosted jet beam of the right hand jet appears to remain largely aligned with the centre of the inflated lobe.

Overall, the pronounced changes in the appearance of the simulated source with changing angle between the jet axis and the line of sight highlight the fact that two radio galaxies with the same underlying physics, such as environment, magnetisation and variability, can appear very different in an observation, purely based on their degree of alignment with our line of sight. In the case of a small angle to the line of sight, the differential Doppler boosting makes the alignment clear. In the case of larger $\theta$, it may not be quite so clear whether one side of a jet is brighter due to Doppler boosting, or due to another process, such as differences in the cluster environment which the jet moves into. To compare simulations and observations it is therefore crucial to build understanding of the range of appearances an individual jet with a specific set of physical characteristics could have based on our viewing angle. This motivates the production of the library of images being released with this work for the purposes of facilitating comparison between the simulations and observations.

Further to pure viewing angle effects, we note that the effects of light travel time on the interpretation of images of out of plane radio sources has been highlighted previously by many authors~\citep[e.g.][]{2002SadunHerculesNucleus}. It would take light approximately 0.4 Myr to travel the end-to-end length of our simulated jets (when reflected to create the double lobe structure) and therefore in observations, we would see the approaching jet at a later time in its evolution than the receding jet. This effect is significant even at a large angle between the jet axis and the line of sight of $60^{\circ}$ -- the light from the tip of the receding lobe still has to travel further to reach us than light from the tip of the approaching lobe by a distance of half of the end to end physical length of the radio lobes. The differential Doppler boosting at $\theta\gtrsim60^{\circ}$ may not be wholly obvious in an observation, such that this may be the range of angles where light travel time effects may complicate the interpretation of observations most strongly.
\subsection{Comparisons to observations of specific sources}
\label{sec:discussion_comparison_obs}
Some of the features seen in our simulations, e.g. large scale asymmetries such as misaligned jets and one-sided backflows, broken jet morphologies and filamentary structures are similar in appearance to features present in observed sources. Here we make some specific comparisons -- we note that the aim of these comparisons is not to definitively identify the causes of behaviour in the specific sources. Instead we aim to illustrate that such a comparison may help to build intuition for how some of the parameters we explore in our simulations may affect emission from radio galaxies and the observations we make of them.
\subsubsection{Large scale asymmetry}
We discussed two kinds of large-scale symmetric behaviour apparent in our highly magnetised simulations in Section~\ref{sec:results_asymmetries}: jets that move away from the central axis and appear misaligned with respect to the centre of the cocoon, and large scale asymmetries in the cocoon shape near to the jet head. 

\cite{2025HortonComplexDR2} describe morphological features of radio galaxies in the second data release of the LOw Frequency ARray Two-metre Sky Survey (LoTSS DR2~\citealt{2022ShimwellLoTSSDR2}), including features such as a misaligned jet axis and whether the jet is continuous, straight or appears restarted. A significant fraction of the sources ($\approx13\%$) were classed as misaligned, with misalignment suggested to be an indicator of precession. Well-resolved examples of misaligned jets can be seen in observations of  Cygnus A~\citep{1989CarilliBroadCygnusA}, 3C 353~\citep{1996SwainJets3C353,1998SwainInternal3C353} and 3C 98~\citep{1997LeahyStudyCM}. 


In particular, the Northern jet of 3C 98~\citep{1997LeahyStudyCM} shares interesting features with our high magnetisation jet with constant energy injection at 3.0 Myr (see Figure~\ref{fig:time_comparison_010}). The Northern jet of 3C 98 appears to curve slightly just before reaching the brightest area at the jet head. The region it curves into is extended away from the jet base on small scales and is misaligned with the current angle of the jet. The small-scale extension of the jet outside of the overall cocoon shape hints that the ram pressure was focused on this area recently, yet the current direction of the beam to large distances is misaligned with this extended region. At 3.0 Myr, our high magnetisation jet with constant energy injection shows a similar focused region of extension at the jet head, with the asymmetric lobe shape caused by recent movement of the jet beam away from the central axis (see Figure~\ref{fig:Bx1_constant_010}).

Precession is often cited as a mechanism for producing multiple hotspots via the dentist's drill effect. \cite{1982ScheuerMorphologySources} suggest that sources with a hotspot that appears further back than the extent of the visible lobe, for example 3C219~\citep{1980PerleyHighGHz}, may be caused by the beam having changed direction such that it is currently excavating a region away from the end and to one side of the cavity. Furthermore, simulations of jet precession have been shown to be able to reproduce multiple hotspots and misaligned jets~\citep[e.g.][]{1990HardeeAsymmetricJet,1991CoxThreeSources,2023HortonNewJets}. Our simulations show that a jet beam can move off axis without being caused by disk precession (as seen in previous RMHD simulations e.g.~\citealt{Mukherjee2020SimulatingDynamics, Meenakshi2023PolarizationFields}). Furthermore we have shown that such movement is likely to have very similar observational signatures to jet beam precession.

\subsubsection{Broken jet morphology}
IC 4296~\citep{1977GossIC4296Galaxy} is an example of a radio galaxy containing a discontinuous jet beam, and can be seen in the radio image made at 1.28 GHz using the MeerKAT telescope~\citep{2021CondonThreads4296}. In the North-Western lobe there is bright emission towards the jet head, forming a large structure, which is nebulous in appearance. This structure appears somewhat separated from the jet base for a section of the jet beam, and just behind this separation (towards the jet base) is a bent region of emission resembling a kinked structure. This feature of the morphology is reminiscent of some epochs of our simulation of a highly magnetised jet with variable energy flux, particularly at 2.5 Myr (see Figure~\ref{fig:seed_6_mag_010_compilation}). We argued that such features which appear helical are likely to be a short-lived phase of the development of the morphology in Section~\ref{sec:appearance_broken_structures} and thus are likely to be seen only in a few sources, such that even a few examples of this behaviour amongst an observed population of sources could have significant implications for the identification of possible jet power variability on $\approx0.1$ Myr timescales.
\subsubsection{Filamentary structures}
Sources observed with high spatial resolution, such as Cygnus A~\citep{1989CarilliBroadCygnusA} and 3C 353~\citep{1996SwainJets3C353,1998SwainInternal3C353} and IC 4296~\citep{1977GossIC4296Galaxy,2021CondonThreads4296}, show filamentary structures spanning tens to hundreds of kpc. Smooth, long filamentary structures are best reproduced in our simulations by the high magnetisation jets, where the flow has some protection from hydrodynamic instabilities. 
\section{Conclusions}
\label{sec:conclusions}
We have simulated four AGN jets in RMHD using the PLUTO code, covering constant and variable injected powers together with high and low magnetisations. These jets evolve over a period of 5 Myr up to lengths of approximately 50 kpc. The variable magnetised jet shows key differences in dynamics, stability and morphology compared to the simulations with low magnetisation and/or constant power.

We summarise our conclusions relating to the interaction between variability and high magnetisation as follows (and as also summarised in Table~\ref{tab:sim_results}):
\begin{enumerate}
    \item Our high magnetisation jets move away from the central axis, resulting in misaligned morphologies.
    \item The constant power high magnetisation jet spends extended periods $\approx 1$ Myr away from the central axis, causing significant asymmetry in the cocoon shape near to the jet head.
    \item Variability in injected jet power prevents the off-axis behaviour from persisting for extended periods of time, such that the cocoon shape remains symmetric on large length scales.
    \item The variable power, high magnetisation jet develops kinked structures over length scales of a few kpc, leading to breaks in the jet material and broken morphologies in simulated radio images.
\end{enumerate}

Additionally, and in agreement with previous works~\citep[e.g.][]{2022LopexMiralles3DBinaries,2024RossiDifferentFlows}, we find that high magnetisation jets with toroidal magnetic fields are more protected from hydrodynamic instabilities, leading to long range filamentary structures and a more clearly defined jet beam to large distances. As expected, the behaviour of the low magnetisation jets is largely reminiscent of that of the non-magnetised jets we presented in~\cite{2026ElleySimulatingBrightening}.

Viewing angle effects have significant consequences on the appearance of all of the above features in simulated radio images. Simulations provide the unique opportunity to watch a single source evolve over time and to rotate it to see it from multiple angles. To illustrate this, and to facilitate the comparison between our simulations and observations of real sources, we have created a library of simulated radio images from different angles for our 4 simulations at different times in their evolution.

\section*{Acknowledgements}
JHM and ELE acknowledge funding from a Royal Society University Research Fellowship (URF\textbackslash R1\textbackslash221062). A.~J.~C. acknowledges support from the Oxford Hintze Centre for Astrophysical Surveys which is funded through generous support from the Hintze Family Charitable Foundation. HW acknowledges funding by the Science and Technology Facilities Council Grant Number ST/W000903/1 and the ISTA-Fellow programme. The authors acknowledge the use of resources provided by the Isambard 3 Tier-2 HPC Facility. Isambard 3 is hosted by the University of Bristol and operated by the GW4 Alliance (https://gw4.ac.uk) and is funded by UK Research and Innovation; and the Engineering and Physical Sciences Research Council [EP/X039137/1]. The authors would like to acknowledge the use of the University of Oxford Advanced Research Computing (ARC) facility in carrying out this work. https://doi.org/10.5281/zenodo.22558. We are grateful for the use of the following software packages: PLUTO~\citep{Mignone2007PLUTO:Astrophysics}, matplotlib~\citep{Hunter2007Matplotlib:Environment}. We are grateful to Dipanjan Mukherjee, Imogen Whittam, Matt Jarvis, Judith Croston and Fraser Cowie for useful discussions. The authors wish to thank the anonymous referee for their helpful and constructive comments which have improved the quality of this manuscript.
\section*{Data Availability}
 
The data underlying this article are available from the authors on reasonable request. Furthermore, the data for the simulated radio images from various angles as discussed in Section~\ref{sec:rendering} are publicly available as FITS files from https://doi.org/10.5281/zenodo.20140021.
\bibliographystyle{mnras}
\bibliography{references}
\appendix
\section{Cartesian versions of magnetic fields}
\label{sec:cartesian_appendix}
In Cartesian coordinates, the magnetic field for $r\leq a$ (see Equation~\ref{eqn:Bfield_within_cylindrical}) is given by
\begin{equation}
    \vec{B}=\begin{pmatrix}
        B_x\\
        B_y\\
        B_z
    \end{pmatrix}=\begin{pmatrix}
        -\frac{B_{\phi, \rm{\,max}}y}{a}\\
        \frac{B_{\phi, \rm{\,max}}x}{a}\\
        0
    \end{pmatrix}.
\label{eqn:Bfield_within_Cartesian}
\end{equation}
Similarly, the magnetic field for $r>a$ is given in Cartesian coordinates by
\begin{equation}
    \vec{B}=\begin{pmatrix}
        B_x\\
        B_y\\
        B_z
    \end{pmatrix}=\begin{pmatrix}
        -\frac{aB_{\phi, \rm{\,max}}}{r^2}y\\
        \frac{aB_{\phi, \rm{\,max}}}{r^2}x\\
        0
    \end{pmatrix}.
\label{eqn:Bfield_outwith_Cartesian}
\end{equation}
For $r>r_j$ we set an initial conditions to a small constant field in a direction perpendicular to the jet given by
\begin{equation}
    \vec{B}=\begin{pmatrix}
        B_x\\
        B_y\\
        B_z
    \end{pmatrix}=\begin{pmatrix}
        0\\
        8.232\times10^{-5}\rm{\,\mu G}\\
        0
    \end{pmatrix}.
\label{eqn:Bfield_ambient_Cartesian}
\end{equation}
Rather than trying to explicitly match these two fields at the boundary, we interpolate between the fields for $r<=r_j$ and $r>r_j$ over a finite distance $\Delta r$, using the same interpolation for both the B and A fields. We interpolate $\vec{A}$ using
\begin{equation}
    \vec{A}(r) = \vec{A}_{\rm{in}}(1-g(r))+\vec{A}_{\rm{out}}g(r),
\end{equation}
and $\vec{B}$ as
\begin{equation}
    \vec{B}(r) = \vec{B}_{\rm{in}}(1-g(r))+\vec{B}_{\rm{out}}g(r),
\end{equation}
where the interpolating function $g(r)$ is given by
\begin{equation}
    g(r) = \frac{1}{2}\left(\tanh{\left(\frac{6(r-r_j)}{\Delta r}-3\right)}+1\right).
\end{equation}
\section{Equilibria from equations of ideal MHD}
We begin from the equations of ideal magnetohydrodynamics (MHD), and assume both a steady state, which is constant in the $z$ direction. We assume negligible resistivity. Working in the frame comoving with the bulk velocity of the jet along the jet direction, the relevant equations of MHD are given by
\begin{equation}
    \vec{\nabla}\cdot\vec{B}'=0,
\end{equation}
\begin{equation}
    \vec{J}'=\vec{\nabla}\times\vec{B}',
\end{equation}
where $\vec{J}'$ is the current and
\begin{equation}
    \rho'\frac{d\vec{v}}{dt}=-\nabla p+\vec{J}'\times\vec{B}'
\end{equation}
which for a steady-state jet can only have a centripetal contribution to the $\frac{d\vec{v}}{dt}$ term, such that
\begin{equation}
    \rho'\frac{v_\phi^2}{r}\hat{r}=-\nabla p+\vec{J}'\times\vec{B}'.
\end{equation}
This leads to the condition for an equilibrium state given by Equation~\ref{eqn:equilibrium_condition}.
\section{Production of simulated radio images}
\label{sec:rendering_appendix}
We give an overview of the method used to produce the simulated images in Section~\ref{sec:rendering}, and expand on further details of the method here. The overall approach is to firstly estimate the rest frame emissivity on a per-cell basis, and then for various lines of sight, Doppler boost this emission (again on a per-cell basis) and sum along the line of sight.

In order to estimate a rest frame emissivity we need the co-moving magnetic field strengths. To calculate these we first split the lab frame magnetic field into two components parallel and perpendicular to the bulk velocity in the lab frame using the following equations
\begin{equation}
    B_\parallel=\frac{B_iv_i}{\sqrt{v_jv_j}},
\end{equation}
such that
\begin{equation}
    B_\perp=\sqrt{B_iB_i-\frac{(B_iv_i)^2}{v_jv_j}}.
\end{equation}
We then apply a Lorentz transformation to the perpendicular component of the magnetic field, assuming no rest frame electric field:
\begin{equation}
    B_\perp'=\frac{1}{\Gamma}B_\perp.
\end{equation}
The parallel component is unchanged ($B_\parallel'=B_\parallel$). We calculate the strength of the rest frame magnetic field as
\begin{equation}
    |B'| = \sqrt{B_\perp'^2+B_\parallel'^2},
\end{equation}
and use this to calculate co-moving emissivities, following
\begin{equation}
    j'(\omega_0')\propto B'^{1.6}u_B'\mathcal{C}\propto B'^{3.6}\mathcal{C},
\end{equation}
as given in the main text (see Equation~\ref{eqn:pseudo_emiss_prop}). To produce the simulated radio images we use the open-source ray tracing code cuDART~\citep{2026WhiteheadcuDARTSources} to Doppler boost the pseudo-emissivity in each cell using its velocity relative to the observer. Here we describe the calculation performed by our rendering code. Frequency transforms as $\omega = D\omega'$, where $D=\Gamma^{-1}(1-\beta\cos\theta)^{-1}$. The emissivity divided by the frequency squared is a Lorentz invariant such that~\citep[e.g.][]{1985RybickiRadiativeAstrophysics}
\begin{equation}
    \frac{j(\omega)}{\omega^2} = \frac{j'(\omega')}{\omega'^2}.
\end{equation}
To calculate the specific intensity in the observer frame we integrate over the line of sight, $\vec{x}=\vec{x_0}+s\hat{s}$:
\begin{equation}
    I(\omega,\vec{x_0},\hat{s})=\int_0^\infty j(\omega,\vec{x})ds
\end{equation}
The rest frame emissivity is a power law such that
\begin{equation}
    j(\omega) = D^2j'(\omega')= D^2j'(\omega_0')\left(\frac{\omega'}{\omega'_0}\right)^{\alpha}.
\end{equation}
Using a telescope operating at a frequency $\omega$, to observe material moving towards the observer, means that the light seen by the telescope was emitted at a lower frequency, and therefore with a higher rest frame emissivity. We therefore transform $\omega'$ to the observer frame to give
\begin{equation}
    I(\omega,\vec{x_0},\hat{s})=\left(\frac{\omega}{\omega'_0}\right)^\alpha\int_0^\infty D(\vec{x})^{2-\alpha}j'(\omega_0',\vec{x})ds.
    \label{eqn:rendering_description}
\end{equation}
where neither $\omega$ or $\omega_0'$ depend on the position along the line of sight and so can be taken outside of the integral. This agrees with the Lorentz invariance of $I_\omega/\omega^3$~\citep[e.g.][]{1985RybickiRadiativeAstrophysics} because we integrate in the observer frame, and as such account for one of the three Doppler factors though the use of $ds$ as opposed to $ds'$. Equation~\ref{eqn:rendering_description} can be summarised as follows:
\begin{enumerate}
    \item Take a single reference value for the emitting frequency $\omega_0'$ and calculate the rest frame specific emissivity at that frequency for each cell.
    \item Multiply each cell's rest frame specific emissivity by its value of $D^{2-\alpha}$ to get the specific emissivity for that cell in the observer frame at $\omega$.
    \item Integrate over the line of sight (in the observer's frame).
\end{enumerate}
Simulated radio images are produced at a resolution half that of the simulation data in each direction due to computational constraints.

\section{Stability of outer jet region}
\label{sec:appendix_stability_outer}
We briefly discuss the stability of the region $a<r\leq r_j$, showing that its dynamics are unlikely to have significant consequences on the dynamics of the jet. Equation~\ref{eqn:force} gives the force on a fluid element slightly displaced from equilibrium. We repeat it here for ease:
\begin{multline}
    \vec{F}(\vec{\xi})=-\vec{\nabla}(-\gamma_{\rm{ad}} p\vec{\nabla}\cdot\vec{\xi}-\vec{\xi}\cdot\vec{\nabla}p)\\-\vec{B}\times(\vec{\nabla}\times(\vec{\nabla}\times(\vec{\xi}\times\vec{B})))+(\vec{\nabla}\times \vec{B})\times(\vec{\nabla}\times(\vec{\xi}\times\vec{B}))
\end{multline}
The magnetic field in the region $a<r\leq r_j$ is given by Equation~\ref{eqn:Bfield_outwith_cylindrical}:
 \begin{equation}
 B_\phi=\frac{a}{r}B_{\phi, \rm{\,max}},
 \end{equation}
 and the pressure in this region is constant (see Section~\ref{sec:methods_pressure_profile}). The first term of Equation~\ref{eqn:force} is therefore zero, as $\vec{\nabla}p=0$ and $\vec{\nabla}(p\vec{\nabla}\cdot\vec{\xi})=0$. Furthermore, $\vec{\nabla}\times\vec{B}=0$, such that the third term of Equation~\ref{eqn:force} also does not contribute. The only remaining term is $-\vec{B}\times(\vec{\nabla}\times(\vec{\nabla}\times(\vec{\xi}\times\vec{B})))$. The displacement is given by
 \begin{equation}
    \vec{\xi}(r,\phi,z',t) = \xi e^{i(m\phi+k'z'-\omega t)}\hat{r}
\end{equation}
The force on the displaced element is given by
\begin{multline}
    -\vec{B}\times(\vec{\nabla}\times(\vec{\nabla}\times(\vec{\xi}\times\vec{B})))\\
    = -\xi\frac{a^2}{4\pi r}B'^2_{\phi, \rm{\,max}}\hat{\phi}\times\left(\vec{\nabla}\times\left(\vec{\nabla}\times\left(\frac{1}{r}e^{i(m\phi+k'z'-\omega t)}\hat{z}'\right)\right)\right)\\
    =-\xi\frac{a^2}{4\pi r}B'^2_{\phi, \rm{\,max}}\hat{\phi}\times\left(-\frac{ik'}{r^2}e^{i(m\phi+k'z'-\omega t)}\hat{r}-\frac{mk'}{r^2}e^{i(m\phi+k'z'-\omega t)}\hat{\phi}\right.\\\left.+\frac{1}{r}\left[-\frac{1}{r^2}e^{i(m\phi+k'z'-\omega t)}+\frac{m^2}{r^2}e^{i(m\phi+k'z'-\omega t)}\right]\hat{z}'\right)
\end{multline}
For $m=1$ this leaves only the imaginary term, which must be zero and so demands $k=0$, so there is no radial force for the $m=1$ mode. For $m=0$, there is a (real) and destabilising force of 
\begin{equation}
    \xi\frac{a^2}{4\pi r^4}B_{\phi, \rm{\,max}}^2e^{i(m\phi+k'z'-\omega t)}\hat{r}= \frac{1}{4\pi r^2}B_\phi^2\vec{\xi}
\end{equation}
which acts to grow or shrink the full column symmetrically. However, the force decreases strongly with increasing $r$, such that the further out $a$ is, the less it is possible for the jet to be affected by this force. Compared to the forces from the inner region $r<a$, given in Equation~\ref{eqn:restoring_force}, this force gets much smaller much quicker with radius. Higher order asymmetric modes $m>1$ will lead to stabilising forces (in the opposite direction to the displacement). Crucially, an increase to the magnetic field strength (by a constant factor across the radius) does not change these conclusions, because there is not the same competition between the thermal and magnetic pressure in this case.
\section{Additional figures}
\begin{figure}
    \centering
    \includegraphics[width=\linewidth]{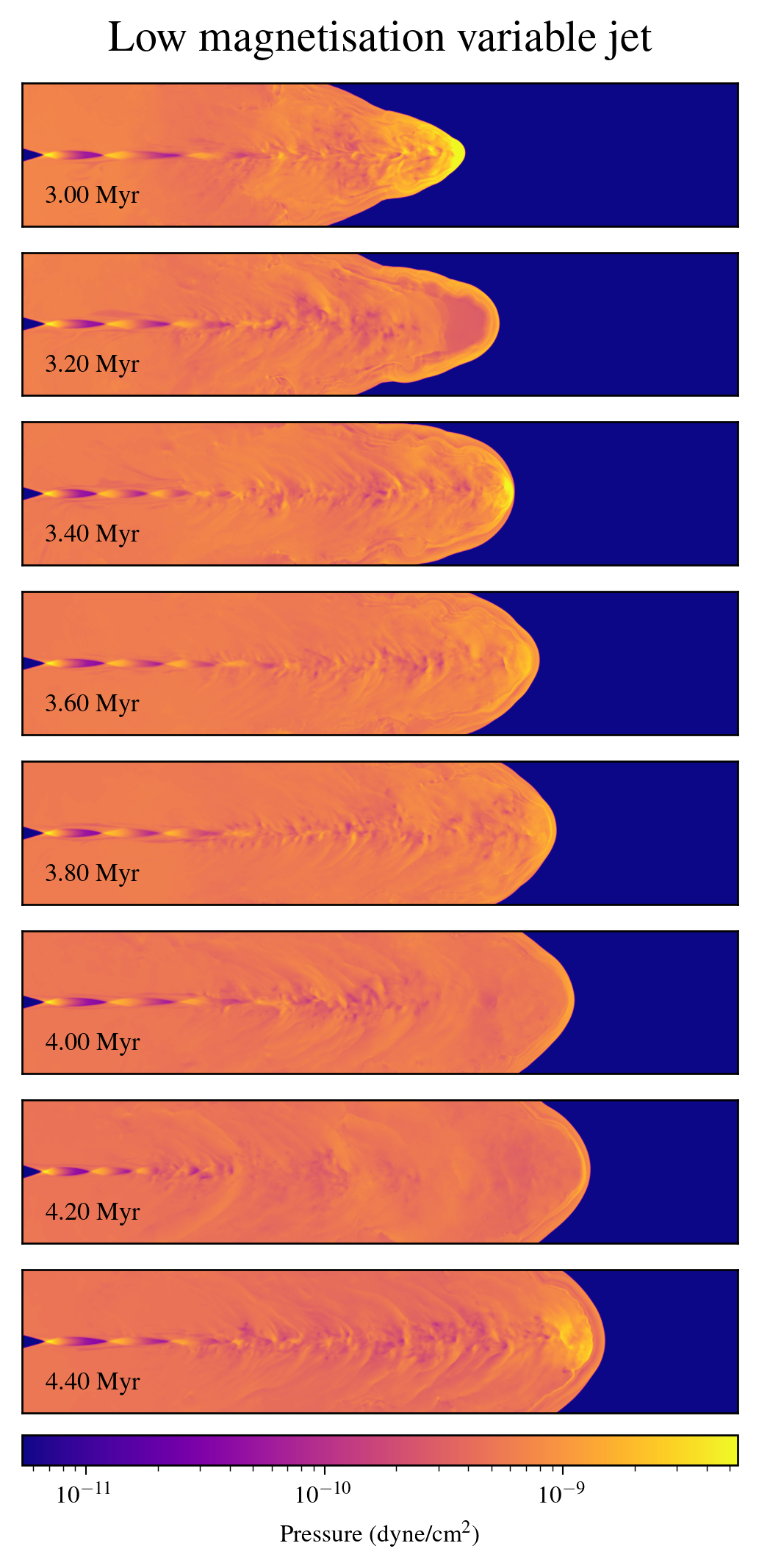}
    \caption{Slices through the variable low magnetisation jet simulation (run V) showing the pressure. The section shown is 60 kpc long. The pressure at the jet head increases and decreases by a factor of approximately $10\times$ during the times shown.}
    \label{fig:changing_hotspot_pressure}
\end{figure}

\begin{figure}
    \centering
    \includegraphics[width=\linewidth]{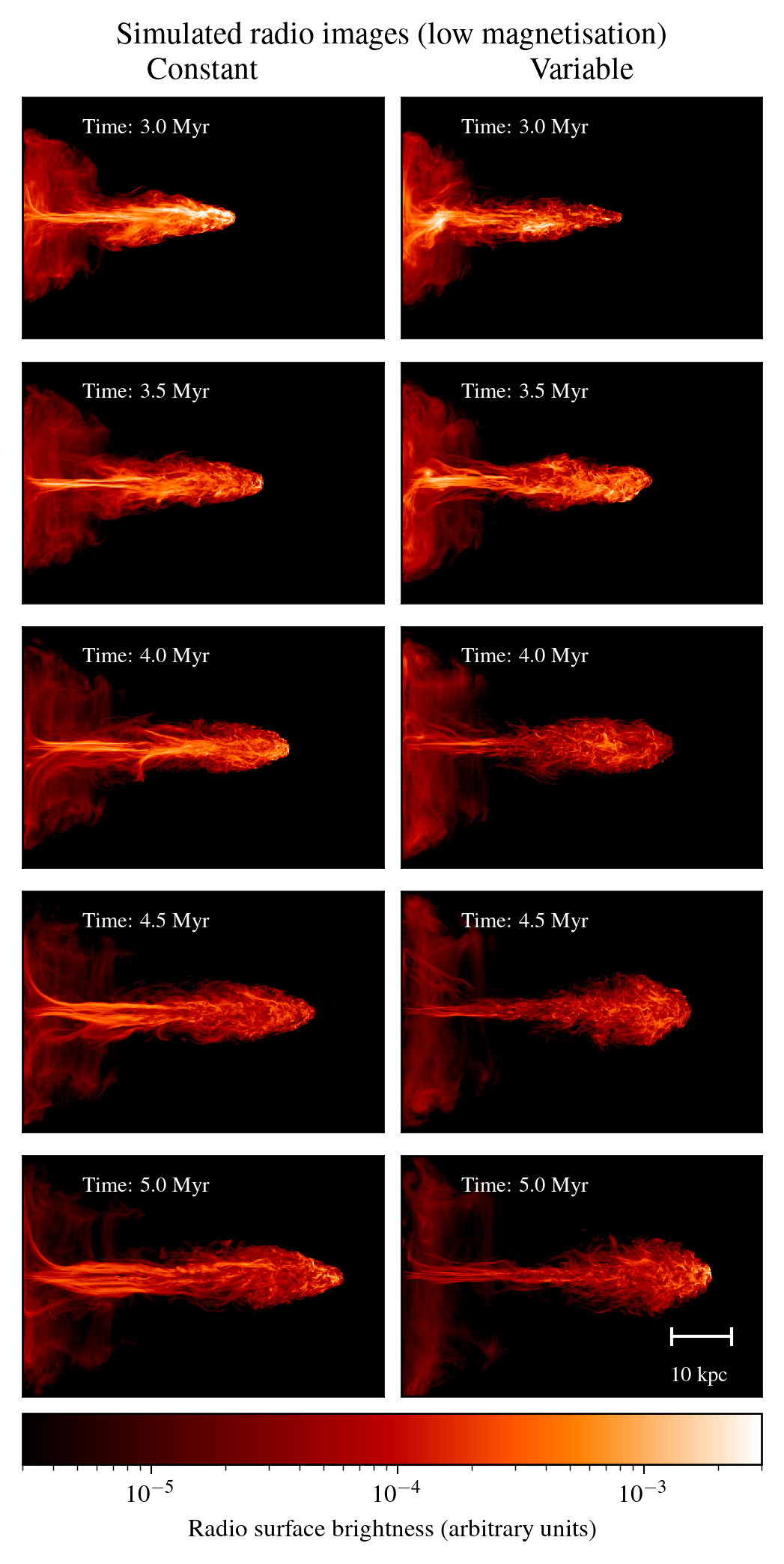}
    \caption{Comparison of simulated radio images of the constant and variable low magnetisation jets at various times in their evolution. This is the low magnetisation counterpart to Figure~\ref{fig:time_comparison_010}. The variable jet shows a bright hotspot at 5.0 Myr, following no bright hotspot at the jet head in either the 4.0 or 4.5 Myr snapshots. The cocoon shape of the constant power jet is similar throughout the evolution, whereas the variable power jet has a thinner cocoon shape at 3.0 Myr (following a period of extended high jet power -- see Figure~\ref{fig:Lorentz_factors_timeseries}) which transitions to a wider shape at later times. This changing cocoon shape can also be seen in the pressure maps in Figure~\ref{fig:changing_hotspot_pressure}. The tracer for the low magnetisation variable jet simulation is shown in Figure~\ref{fig:broken_jets}.}
    \label{fig:time_comparison_001}
\end{figure}

\begin{figure*}
    \centering
    \includegraphics[width=\linewidth]{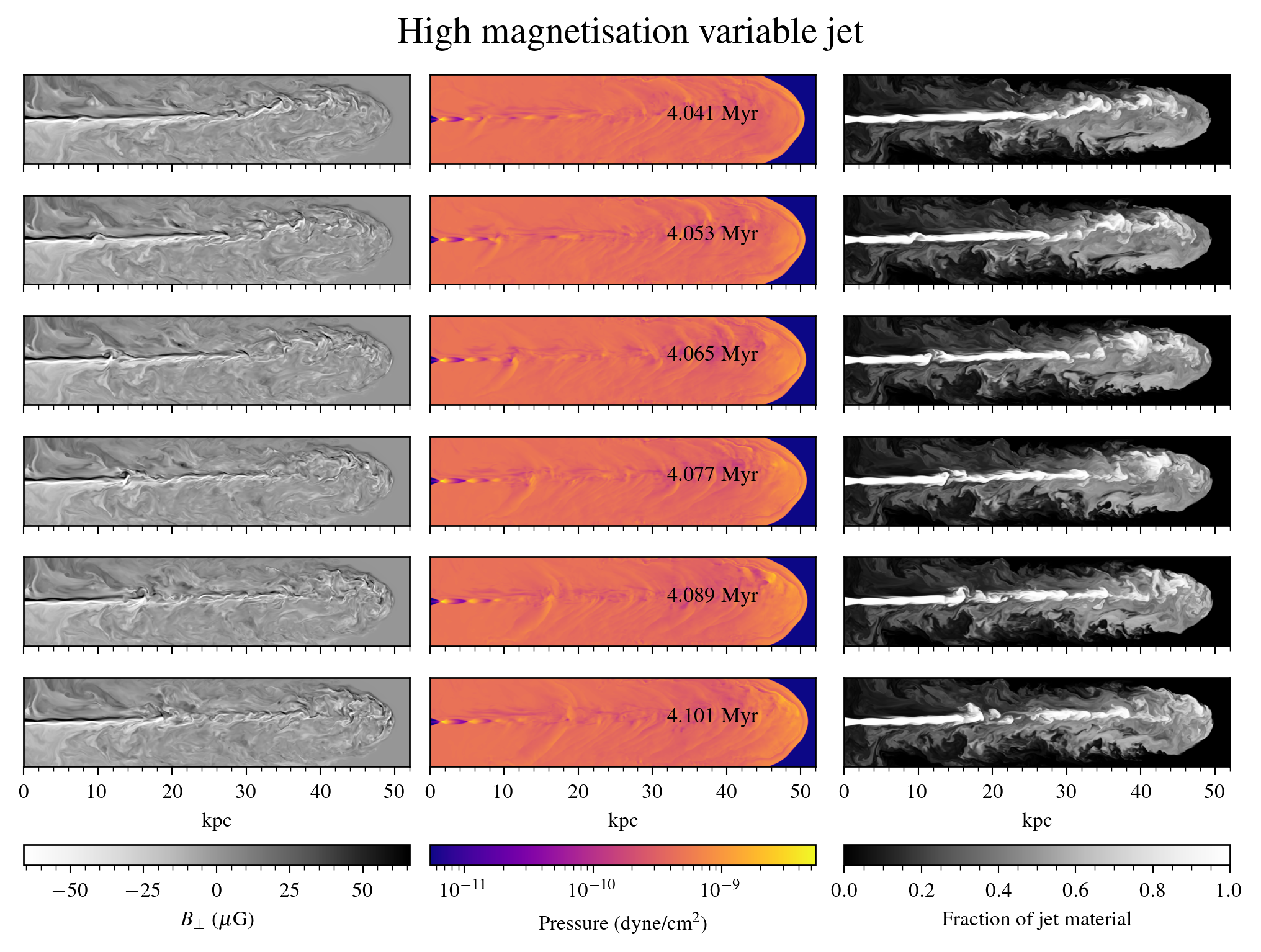}
    \caption{Slices through the variable high magnetisation jet simulation (run VM), showing the component of the magnetic field perpendicular to the slice ($B_\perp$, toroidal component at the jet base), pressure and the fraction of jet material (tracer). Limits on the pressure colour scale are chosen to accentuate the travelling shock structures, and do not accurately reflect the ambient medium pressure.}
    \label{fig:break_growth_following_decrease}
\end{figure*}

\begin{figure*}
    \centering
    \includegraphics[width=\linewidth]{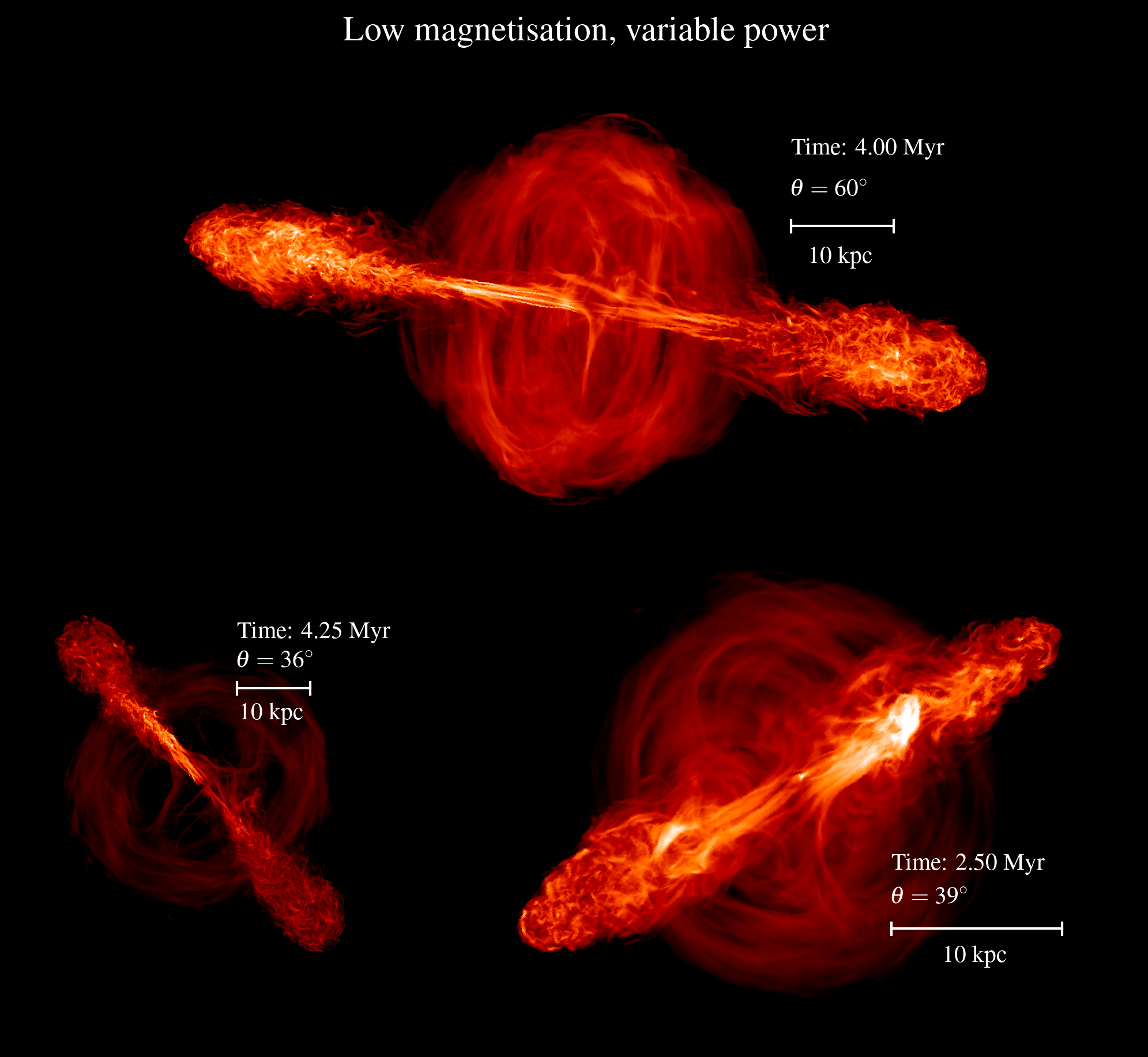}
    \caption{Three illustrative examples of simulated radio images for the low magnetisation, variable power jet simulation at different times in its evolution. See Figure~\ref{fig:seed_6_mag_010_compilation} for the corresponding times and views of the high magnetisation jet simulation. In comparison, the low magnetisation jet simulation shows greater signs of turbulence close to the jet head, whereas the high magnetisation jet simulation has filamentary structures on longer length scales, as discussed in Section~\ref{sec:results_low_mag_variable_mixing}. All images are made with a dynamic range of 3 dex in brightness, however the limits of the brightness scale vary between images to account for differences in overall brightness due to age and boosting. The spatial scale for each image is indicated by the corresponding scale bar; images are presented at different spatial scales to facilitate seeing various details and features in each. $\theta$ gives the angle between the approaching (brighter) jet and the line of sight.}
    \label{fig:seed_6_mag_001_compilation}
\end{figure*}

Here we present some additional figures, for the purposes of comparison, or confirmation of agreement with previous results. Figure~\ref{fig:changing_hotspot_pressure} shows the pressure in the low magnetisation jet, confirming that the hotspot pressure once again changes significantly for periods on the order of approximately 0.1 Myr as in the non-magnetised jets of~\cite{2026ElleySimulatingBrightening}. Figure~\ref{fig:time_comparison_001} is the low magnetisation counterpart to Figure~\ref{fig:time_comparison_010}, showing the evolution of the low magnetisation jet simulations using simulated radio images. Figure~\ref{fig:break_growth_following_decrease} shows the formation of a discontinuity in the jet material as the result of interactions between the magnetic field and a discontinuity in the pressure field following a discrete decrease in the rate of change of jet power (as opposed to Figure~\ref{fig:break_growth_seed6_010}, which shows a similar process as a result of a discrete increase in the rate of change of jet power). Finally, Figure~\ref{fig:seed_6_mag_001_compilation} shows the low magnetisation high power jet simulation at various times and viewing angles and is included here as a comparison to the high magnetisation case shown in Figure~\ref{fig:seed_6_mag_010_compilation}.
\end{document}